\documentclass[fleqn,usenatbib]{mnras}

\usepackage{newtxtext,newtxmath}

\usepackage[T1]{fontenc}

\DeclareRobustCommand{\VAN}[3]{#2}
\let\VANthebibliography\thebibliography
\def\thebibliography{\DeclareRobustCommand{\VAN}[3]{##3}\VANthebibliography}

\usepackage{graphicx}	
\usepackage{amsmath}	

\usepackage{amssymb}	
\usepackage{xspace}
\usepackage[colorinlistoftodos]{todonotes}

\newcommand{\kms}{km\ s$^{-1}$\xspace}

\newcommand{\msun}{M$_\odot$\xspace}	
\newcommand{\rhalf}{R$_\mathrm{50}$\xspace}	
\newcommand{\zhalf}{z$_\mathrm{50}$\xspace}

\title[Disc growth and heating]{Disc growth and vertical heating of lenticular  galaxies in the Fornax cluster}

\author[M. Martig et al.]{Marie Martig$^{1}$\thanks{E-mail: m.martig@ljmu.ac.uk}, Francesca Pinna$^{2,3}$, Jes{\'u}s Falc{\'o}n-Barroso$^{2,3}$, Ignacio Mart{\'i}n-Navarro$^{2,3}$, Ivan Minchev$^{4}$, \newauthor Yuchen Ding$^{1}$
\\
$^{1}$Astrophysics Research Institute, Liverpool John Moores University, 146 Brownlow Hill, Liverpool L3 5RF, UK\\
$^{2}$Instituto de Astrof{\'i}sica de Canarias, Calle Via L{\'a}ctea s/n, 38200 La Laguna, Tenerife, Spain\\
$^{3}$Depto. Astrof{\'i}sica, Universidad de La Laguna, Calle Astrof{\'i}sico Francisco S{\'a}nchez s/n, 38206 La Laguna, Tenerife, Spain\\
$^{4}$Leibniz-Institut f\"{u}r Astrophysik Potsdam (AIP), An der Sternwarte 16, D-14482 Potsdam, Germany\\
}
\date{Accepted XXX. Received YYY; in original form ZZZ}

\pubyear{2025}

\begin{document}
\label{firstpage}
\pagerange{\pageref{firstpage}--\pageref{lastpage}}
\maketitle

\begin{abstract}
We present a detailed analysis of the vertical and radial structure of mono-age stellar populations in three edge-on lenticular galaxies (FCC 153, FCC 170, and FCC 177) in the Fornax cluster, using deep MUSE observations. By measuring the half-mass radius (\rhalf) and half-mass height (\zhalf) across 1 Gyr-wide age bins, we trace the spatial evolution of stellar populations over cosmic time. All galaxies exhibit a remarkably constant disc thickness for all stars younger than $\sim 6$ Gyr, suggesting minimal secular heating and limited impact from environmental processes such as tidal shocking or harassment. Evidence of past mergers (8--10 Gyr ago) is found in the increase of \zhalf for older populations. We find that accreted (metal-poor) stars have been deposited in quite thick configurations, but that the interactions only moderately thickened pre-existing stars in the galaxies, and only caused mild flaring in the outer regions of the discs.  The radial structure of the discs varies across galaxies, but in all cases we find that the radial extent of mono-age populations remains constant or grows over the past 8 Gyr. This leads us to argue that within the radial range we consider, strangulation, rather than ram-pressure stripping, is the dominant quenching mechanism in those galaxies. Our results highlight the usefulness of analysing the structure of mono-age population to uncover the mechanisms driving galaxy evolution, and we anticipate broader insights from the GECKOS survey, studying 36 nearby edge-on disc galaxies.
\end{abstract}




\section{Introduction}
Galactic discs have a complex radial and vertical structure which results both from their cosmological accretion histories and from internal processes.
The radial structure of disc galaxies is often described as resulting from inside-out formation \citep{Larson1976}. In this model, the inner regions of galaxies assemble first, forming discs that are initially small and dense \citep{Mo1998}. Then, as the Universe expands, the angular momentum of the gas falling onto the disc increases, so that this gas can progressively settle at larger and larger galactocentric radius, ultimately leading to a progressive growth of the stellar disc. Simulations have shown that the picture is actually more complex: the evolution of angular momentum in galaxies is driven by the details of cold gas accretion  along filaments \citep{Pichon2011, Stewart2013, Danovich2015}, by galaxy interactions \citep{Lagos2018, Cadiou2022}, and by feedback and galactic fountains redistributing the gas within galaxies \citep{Maller2002,Dutton2009,Brook2012,Grand2019}. In addition to these processes, radial migration can alter the radial distribution of stars in discs, and change the angular momentum and the scale-length of stars after their birth \citep{Sellwood2002,Frankel2019}. In spite of all this, most simulations of Milky Way mass disc galaxies show a remarkably regular radial structure, with younger stars having larger scale-lengths \citep{Bird2013, Stinson2013, Martig2014a, Minchev2015, Buck2020}. Mergers can temporarily alter this relation \citep{Martig2014a}, but, at least in Milky-Way type galaxies, do not affect the overall trend over long timescales.

In observations, this inside-out growth manifests as negative radial age gradients, commonly observed in massive late-type galaxies \citep{SanchezBlazquez2014,Wilkinson2015,Parikh2021, Pessa2023}. This is also evident from studying 
spatially-resolved  star formation histories \citep{CidFernandes2013, GonzalezDelgado2014, Sacchi2019}, or from slicing galaxies in different age bins and showing that older stars are more concentrated towards the centre of galaxies \citep{Peterken2020, Martig2021}. Using observations of resolved stars with the Hubble  Space
Telescope, \cite{Williams2009} and \cite{Gogarten2010} have also directly shown that the scale-length of stellar populations increases for younger stars in M33 and NGC300.  In the Milky Way, measuring the scale-lengths of stars of different ages is a challenging task, but recent studies suggest that younger populations indeed have larger scale-lengths \citep{Amores2017, Xiang2018} and it is well-established that metal-rich, low-[$\alpha$/Fe] stars have a larger scale-length than metal-poor, high-[$\alpha$/Fe] stars which are generally older  \citep{Bensby2011,Bovy2012b, Cheng2012b, Martig2016}.

The vertical structure of discs is also complex. In many simulated galaxies, young populations are found in a thin disc component with a low velocity dispersion, and the scale-height and vertical velocity dispersion of mono-age populations increases with age \citep{House2011, Bird2013, Stinson2013, Martig2014a,Martig2014b, Grand2016, Ma2017, Buck2020, Park2021, Agertz2021}. While this general trend is well established, there is more disagreement on the origin of this relation, and in particular how much of it is due to the birth conditions of stars compared to later dynamical heating.

While a few simulations report the birth of stars in thin discs already at early times
\citep{Meng2021,Tamfal2022}, in many simulated galaxies, stars are intrinsically born in thicker discs at high redshift. This might be due to an early period of gas-rich mergers \citep{Brook2004}, but also to high levels of turbulence in gas rich discs \citep{Bournaud2009, Ceverino2017}, with high levels of feedback from supernovae and bursty star formation \citep{Yu2021}.
As time progresses, the decline in gas fraction and gravitational disc instability, together with a decrease in star formation rate and feedback, allow the formation of progressively thinner discs \citep{Bird2013,Kassin2014, Grand2016, Ma2017, Ceverino2017, Dubois2021, vanDonkelaar2022}. \cite{Stern2021} and \cite{Yu2021} also propose that the virialisation of the inner circumgalactic medium (when haloes reach a typical mass of $\sim 10^{12}$ \msun) confines the gas discs and allows the formation of thin stellar discs. 

After birth, stars also undergo some later dynamical heating, via mergers and secular processes \citep{Quinn1993, Villalobos2008,Martig2014b, Ma2017, Buck2020, Bird2021, Agertz2021, Park2021, Yu2023}. How much of a stellar population's thickness reflects its birth properties vs later heating varies with time \citep{McCluskey2024}, and also varies from galaxy to galaxy, depending on each galaxy's history \citep{Pinna2018} and mass \citep{Leaman2017}.

The smoothness of the transition from thick to thin populations can be assessed by studying the scale-height (or velocity dispersion) of stars as a function of their age. 
In particular, mergers can create sharp vertical jumps in the relation between age and velocity dispersion or thickness: this happens when populations born before the merger are thickened, while after the merger, gas cools down again, and younger populations are formed with a distinctly lower thickness \citep{Quillen2001,House2011, Martig2014b, Grand2016}.

The assembly history of a galaxy is not just encoded in the average scale-heights of mono-age populations, but also in how these scale-heights vary with radius. Simulated galaxies show a large diversity in the radial profiles of their scale-heights, from cases where mono-age populations do not flare, to cases where they all flare strongly, and cases where either the old or young populations flare the most \citep{Bird2013, Minchev2015, Grand2017, Ma2017, Buck2020, Agertz2021, Garcia2021, Sotillo2023}. This does not necessarily translate into flaring for the overall stellar disc, which can be flat even when all mono-age  populations  flare \citep{Minchev2015, Garcia2021}. Flaring can in some cases already be imprinted at birth if the gas disc itself is flared \citep{Bird2013, Grand2017}, but can also be due to secular evolution \citep{Minchev2012}, or mergers \citep{Bournaud2009, Qu2011}.  In detail, \cite{Sotillo2023} propose that the relation between merger history and flaring is complicated, with a weak indication that galaxies with recent major mergers tend to have young populations that are more flared than older populations.

Observationally, the vertical structure of the disc has been extensively studied in the Milky Way.  It has been shown that metal-poor, $\alpha$-enhanced (i.e., older) stars have larger scale-heights than metal-rich, $\alpha$-poor (i.e., younger) stars \citep{Bovy2012b}. More recently, \cite{Mackereth2017} and  \cite{Xiang2018} have also directly shown that scale-height increases with age. Similarly, the vertical velocity dispersion of stars increases with age \citep{Quillen2001, Soubiran2008, Sanders2018, Anders2023}. This results in an increase of the average age of stars as a function of height above the midplane \citep{Casagrande2016, Martig2016}.
In terms of radial structure, the geometric thick disc of the Milky Way has a negative radial age gradient \citep{Martig2016, Imig2023, Anders2023}, which is consistent with being formed from the superposition of flared mono-age populations (as proposed by \citealp{Minchev2015}). Direct measurements of the radial profiles of mono-age population later showed that they do indeed flare. The strength of flaring as a function of age is debated: while early papers showed that the youngest populations are the  most  strongly  flared \citep{Bovy2016, Mackereth2017}, more recent works seem to show the opposite  \citep{Imig2025, Khoperskov2025}, possibly because of the different radial coverage of the data used.

In external galaxies,  thick discs usually have older populations than thin discs \citep{Mould2005, Rejkuba2009,Yoachim2008b,Comeron2015,Comeron2016, Kasparova2016,Kasparova2020, Eigenbrot2018, Pinna2019a,Pinna2019b,Scott2021, Martig2021, Sattler2023, Sattler2025}, but detailed measurements of the thickness or velocity dispersion of stars as a function of age are rare.
In M31, observations of individual stars \citep{Dorman2015} and planetary nebulae \citep{Bhattacharya2019} show that the line-of-sight velocity dispersion increases as a function of age. This is also found in M33 using observations of star clusters \citep{Beasley2015}. For edge-on galaxies slightly further away, the properties of resolved stars can be studied with deep Hubble Space Telescope images. These show that older stars have larger scale-heights \citep{Seth2005, Tikhonov2005}, and for 3 low mass galaxies, \cite{Streich2016} show that the youngest stars have a constant scale-height as a function of radius, while older populations exhibit a mild flaring.

For galaxies even further away, spectroscopy is the only way to determine the age structure of discs, but this requires deep observations to reach adequate signal-to-noise ratios in the thick disc-dominated regions. \cite{Poci2019, Poci2021} used such deep observations, and with the help of dynamical modelling measured the age-velocity-dispersion relation for four galaxies in total, showing an increase of velocity dispersion with age. A few other papers \citep{Guerou2016,Pinna2019a,Pinna2019b,Martig2021} show maps of stars of different ages, where it is apparent that younger stars are in thinner (sometimes possibly flared) discs, but this effect has not been quantified.

\begin{table*}
\caption{Main properties of FCC 153, FCC 170 and FCC 177: distance  \citep{Blakeslee2009}, effective radius in the $r$-band \citep{Iodice2019b}, optical radius in the $B$-band \citep{deVaucouleurs1991}, stellar mass \citep{Iodice2019b}, maximum circular velocity \citep{Bedregal2006}}
\label{tab:properties}
\begin{tabular}{lccccccc}
\hline
 & $D$ (Mpc)  & $R_{e}$ (")& $R_{e}$ (kpc) & $R_{25}$ (") & $R_{25}$ (kpc)  & $M_*$ ($10^{9}$ \msun) & $V_c$ (\kms)\\
 \hline
FCC153 &20.8 & 19.8&2.0 & 77.1&7.8 & 7.6 & 165\\
FCC170 & 21.9& 15.9&1.7 & 80.5&8.6 & 22.5& 280\\
FCC177 &20.0 & 35.9 &3.5& 72&6.9  &  8.5& 120\\
\hline
\end{tabular}
\end{table*}
\cite{MartinNavarro2019} showed how Integral Field Spectroscopy (IFS) can be used to quantify the morphology of stellar populations of different ages, and applied this approach to massive early-type galaxies ("red nuggets"). In this paper, we present the first detailed analysis of the radial and vertical structure of mono-age populations in external disc galaxies, using IFS data from the Multi Unit Spectroscopic Explorer \citep[MUSE,][]{Bacon2010} on the Very
Large Telescope. We focus on three lenticular galaxies   first presented in \cite{Pinna2019a,Pinna2019b}, and also studied in  \cite{Poci2021} and \cite{Ding2023} with dynamical models.
These galaxies are ideal for this first analysis as they are now well studied, and they  have a regular structure, with a mid-plane unobscured by dust and no emission lines complicating the analysis of spectra.  The MUSE observations were obtained by the Fornax3D survey \citep{Sarzi2018}, a survey of the 33 brightest galaxies within the virial radius of the Fornax cluster. At a distance of $\sim$ 20 Mpc, Fornax has a total mass of $7 \times 10^{13}$ \msun with a projected radius of 1.4 Mpc  \citep{Drinkwater2001}. The high density of galaxies in the core of the cluster, and the hot intracluster gas \citep[detected in X-rays,][]{Su2017} both affect the evolution of galaxies in the cluster \citep{Iodice2019a, Spavone2022}: we do not necessarily expect them to follow the relations between age, scale-height and scale-length observed in Milky Way-like galaxies. In particular, they might not have formed inside out: stars in their inner disc are typically younger than in the outer disc \citep{Bedregal2011, Ding2023}.  Stellar discs might also be partially disrupted by galaxy harassment \citep{Moore1998} and tidal shocking \citep{Joshi2020, Ding2024}, which could affect their vertical structure. In this paper, we explore the possible impact of those processes on the detailed structure of mono-age populations.

In Section \ref{sec:galaxies} we introduce the three galaxies studied in this paper. In Section \ref{sec:method} we then describe the data, the method used to extract the properties of the stellar populations, and the way we quantify the radial and vertical structure of the galaxies. In Section \ref{sec:results} we present our results on the radial and vertical structure of mono-age populations, and discuss our findings and the limitations of our analysis in Section \ref{sec:discussion}.

\section{Galaxy properties}\label{sec:galaxies}

We study three S0 galaxies that are part of the Fornax3D survey \citep{Sarzi2018}. Their kinematics, stellar populations and formation histories have been studied in detail by \cite{Pinna2019a} for FCC 170 and \cite{Pinna2019b} for
FCC 153 and 177. The main properties of the galaxies are listed in Table \ref{tab:properties}.

FCC170 (NGC 1381) is the most massive of the three galaxies. In its centre, it  possibly hosts a classical bulge \citep{Williams2011} but the dominant component is a boxy, X-shaped bulge \citep{Lutticke2000, Bureau2006} indicating the presence of a bar. A small nuclear disc is visible in the kinematics maps presented in \citep{Pinna2019a}, and a nuclear star cluster was detected by \cite{Turner2012}. The stellar disc is well fitted by the superposition of a thin and a thick component \citep{Comeron2018}.
In spite of this morphological complexity, the stellar populations in FCC170 are relatively simple. The spectra do not show any emission lines \citep{Williams2011}, and the stellar age map presented by \cite{Pinna2019a} is very uniform: most of the stars in the galaxy are very old, with very few regions younger than 8 Gyr.  From a detailed study of the stellar populations in FCC170, \cite{Pinna2019a} highlighted the presence of stars that are younger and more metal-poor than the bulk of the galaxy. They propose that these stars were accreted, and that FCC 170 had a merger with a $\sim 2.5 \times 10^9$\msun galaxy, possibly 10 Gyr ago \citep[see also][]{Galan2023}. The overall evolution of this galaxy was also probably strongly affected by its large-scale environment, as it is currently within a high-density region of the Fornax cluster (the north-south clump, \citealp{Iodice2019a}).

The other galaxies, FCC 153 (IC 335) and FCC 177 (NGC 1380A) are about three times less massive than FCC 170. Their bulges were classified as elliptical (for FCC 153) and as boxy (for FCC 177) by \cite{Lutticke2000}, with no signs of nuclear stellar discs. However, both galaxies host a blue nuclear star cluster \citep{Turner2012}. Their stellar discs can both be decomposed into a thin and a thick component \citep{Comeron2018}. For both galaxies, \citet{Pinna2019b} report the presence of accreted stars, contributing about 4 per cent of the total stellar mass. The main difference between these two galaxies and FCC 170 is that they have extended star formation histories, and a thin disc that is significantly younger than the thick disc \citep{Pinna2019b}.  This is possibly related to the fact that they are currently in lower density regions of the Fornax cluster compared to FCC 170, with FCC 153 being the furthest away from the cluster core \citep{Iodice2019a}.  They also probably fell into the Fornax cluster at a later time compared to FCC 170 \citep{Ding2023}.

Overall, these three galaxies represent the diversity of populations of lenticular galaxies in Fornax. This paper aims to further explore this diversity by studying the radial and vertical structure of their mono-age populations, to shed some light on their formation history of their discs.

\section{Data and analysis} \label{sec:method} 
\subsection{Observations and data reduction}
We use data from the Fornax3D survey \citep{Sarzi2018}, which used MUSE \citep{Bacon2010} on the Very Large Telescope.  The observation strategy, data quality and position of pointings are described in \cite{Sarzi2018}. Each galaxy is covered by two pointings: the central one has an on-source time
of one hour for all galaxies, and the outer one has an on-source time of 1.5 hour for FCC 153 and FCC 177, and 2 hours for FCC 170.

Data reduction was performed using the MUSE reduction pipeline version 1.6.2 \citep{Weilbacher2016,Weilbacher2020} and the pointings were stitched together, as described in detail in \cite{Sarzi2018} and \cite{Pinna2019a}. The mosaics were released in 2023\footnote{\href{https://www.eso.org/sci/publications/announcements/sciann17540.html}{https://www.eso.org/sci/publications/announcements/sciann17540.html}}, and are publicly available from the ESO archive.

 \subsection{Stellar populations analysis}
We follow the general steps described in \cite{Pinna2019a}. This starts with performing a Voronoi binning of the combined cube \citep{Cappellari2003}, aiming for a signal-to-noise ratio of 60 per spatial bin in the range from 4750 to 5500 \AA \ (including all spaxels with an individual signal-to-noise ratio above 1). This results in 4863 bins for FCC 170, 3583 bins for FCC 153 and 3213 bins for FCC 177. Of those, we later discard spectra for which no good fit can be obtained, or which were contaminated by foreground stars. 

The extraction of stellar populations properties for the galaxies is simplified by the absence of emission lines in their spectra. This means that we can simply fit the spectra using the Penalized Pixel-Fitting code described in \cite{Cappellari2004} and \cite{Cappellari2017}, without masking or subtracting emission lines first. We fit the stellar kinematics and populations at the same time, using 8$^\mathrm{th}$ order multiplicative polynomials to account for
uncertainties in the spectral calibration.

To fit the spectra, we use the MILES single stellar population models (SSPs) in their [$\alpha$/Fe]-variable version  based on BaSTI isochrones \citep{Vazdekis2015}, assuming a Kroupa Universal IMF, with a slope of 
1.30. We restrict ourselves to a range of ages and metallicities which are deemed safe: we include values of total metallicity [M/H] from -1.79 to 0.4  dex, and  values of age between
0.1 and 14.0 Gyr. [$\alpha$/Fe] can only take two values, either 0.0 or 0.4 dex.
We fit the spectra  from 4750 to 5500 \AA~using this set of SSPs, using a regularisation parameter of 0.5.

To estimate the uncertainties in stellar population properties, after this first fit we perform 50 Monte Carlo iterations. For each iteration, we add Gaussian noise to each spectrum. For this, we first compute the difference between each spectrum and its best fit, and measure the standard deviation of those residuals. Then we draw a vector of random numbers following a Gaussian distribution centred on zero, and with a standard deviation equal to the standard deviation of the residuals. This noise vector is added to the initial spectrum, and the new spectrum is fitted using the exact same procedure as described previously. This process is repeated 50 times for each Voronoi bin.

\subsection{Maps of mono-age populations}

\begin{figure*}
\includegraphics[width=\textwidth]{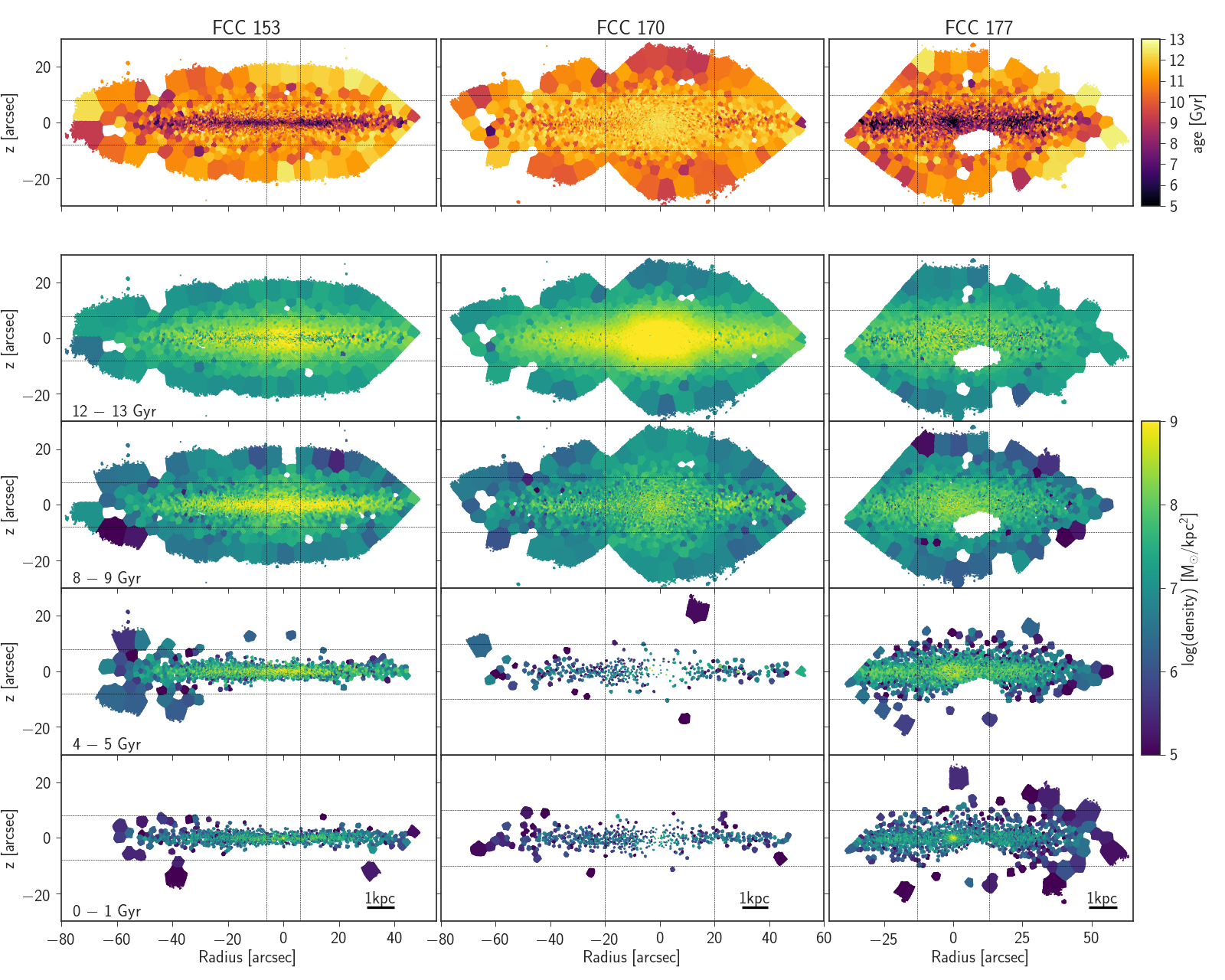}
\caption[]{The top row shows the average mass-weighted age distributions for the three galaxies studied here, while the rest of the figure shows the surface density for four different mono-age populations in each galaxy (12--13 Gyr old, 8--9 Gyr old, 4--5 Gyr old, and finally 0--1 Gyr old). In all panels, the vertical lines enclose the central bulge or bar-dominated region, while the horizontal lines delimitate the thin- and thick disc-dominated regions \citep{Pinna2019a,Pinna2019b}. In all galaxies, we find strong differences in the spatial distribution of stars of different ages: one striking aspect is the much thinner distribution of young stars compared to older stars.} 
\label{fig:maps_all}
\end{figure*}
We aim to study the spatial distribution of mono-age populations, which we here choose to be populations within age bins that are 1~Gyr wide, from 0 to 14~Gyr. This requires to create maps of the mass distribution for each mono-age population. To do this, we  need to estimate the total stellar mass in each spaxel, and then compute the fraction of that mass that belongs to each mono-age population.

The total mass in each spaxel is computed following the exact same procedure as described in \cite{Pinna2019a}. In summary, this involves first computing the absolute \textit{V}-band surface brightness for each spaxel, by converting a \textit{g}-band image from the Fornax Deep Survey \citep{Iodice2019b} to the \textit{V}-band and comparing it to our MUSE data. Then, we compute the mass-to-light ratio of each Voronoi bin from the best fitting combination of SSPs reproducing its spectrum, and assign this mass-to-light ratio to all spaxels within a Voronoi bin. Finally, the mass in each spaxel is given by Equation 1 in \cite{Cebrian2014}, using their absolute \textit{V}-band surface brightness and mass-to-light ratio.

The mass weights from pPXF can then be used to give for each Voronoi bin the fraction of the total mass in a given age range (summing over all possible values of [M/H] and [$\alpha$/Fe]). By multiplying this fraction with the total mass in each spaxel, we finally obtain the mass belonging to each mono-age population. From this, we can build maps such as the ones shown in Fig. \ref{fig:maps_all}: there we show surface density maps for each galaxy for 4 representative mono-age populations, from old to young. For each galaxy, we have 14 such surface density maps (see Appendix \ref{sec:appendix_maps} for all these maps), which we then use to quantify the spatial structure of mono-age populations.

We perform this whole process not just for the best fit from pPXF, but also for each of our 50 Monte Carlo iterations, so that we can estimate the uncertainties in our recovery of the spatial distribution of the mono-age populations.

\subsection{Quantification of the radial structure}
From the surface density maps of each mono-age population, we first 
quantify the radial structure of our three galaxies. This could be done by fitting the maps with a sum of parametrized components representing a disc, a bulge, a nuclear disc, or any other necessary component. However, this would be a complex procedure, particularly in the case of noisy data for populations that are not very massive. Instead, we choose a much simpler approach which is to measure the half-mass radius, $R_{50}$ (the radius enclosing half of the mass). We compute both a global value of $R_{50}$ for the whole galaxy, and a value for each mono-age population. We note that these are not the true half-mass radii of the galaxies, as we do not observe the whole extent of the discs. However, the values we measure are still useful to compare populations within a given galaxy.

We determine $R_{50}$ by computing a cumulative mass profile for all the individual spaxels, using their distance from the minor axis of the galaxies (we call this distance "radius" from now on). 
The pointings do not cover each side of the galaxies symmetrically, and in particular on one side of the galaxies the shape of the field-of-view restricts the range of vertical distances covered by the data. To ensure that this does not bias our analysis, we restrict ourselves to regions of the galaxies that are fully covered by the MUSE field-of view ($x< 30$ arcsec for FCC 153 and FCC 170, and $x>-10$ arcsec for FCC 177).
This leaves us with significant differences in the extent of the radial coverage of the two sides of the galaxies. One solution would be to use only the side with the most extended coverage to compute \rhalf. Instead, to improve our statistical uncertainties, we symmetrize the data based on the most complete side: we assume that, on the side with the narrowest radial coverage, the mass profile in the region not covered by the pointing is the same as on the other side of the galaxy. This allows to build a cumulative mass profile using all the information available to us.

We follow the same procedure to compute $R_{50}$ for mono-age populations, using the individual mass maps created for each population, and use our 50 Monte Carlo iterations to estimate an uncertainty on those values.

\subsection{Quantification of the vertical structure}

The vertical structure of mono-age populations is often analysed by fitting the vertical density profiles by exponential or sech$^2$ functions to measure a scale-height, or by measuring the half-mass height \zhalf. Given that those measurements are often equivalent \citep[e.g.,][]{Sotillo2023}, we choose here to use  the half-mass height \zhalf: this is easier to compute and less model-dependent than scale-heights, and is meaningful even if density profiles do not follow a perfect exponential or sech$^2$ shape.

To study how the thickness of the disc varies as a function of radius, we compute radial profiles of \zhalf for the galaxy as a whole and for mono-age populations. We define radial bins that are within a given distance of the minor axis, up to 65 arcsec in distance for FCC~153 and FCC 170, and up to 55 arcsec for FCC 177. Each bin has a width of 5 arcsec: we find that this is a good compromise, as smaller bins give noisier profiles and larger bins erase some of the interesting structures. Within each radial bin, we compute a cumulative vertical mass profile, which allows us to directly determine \zhalf for that radius. We do this both for the total mass distribution, and for all mono-age populations. 

We also compute an average \zhalf for each mono-age population, as an indicator of disc thickness for that age. This is meaningful because, as we will later show, there is a region in each galaxy where the \zhalf of mono-age populations does not vary much with radius: we exclude both the inner galaxy (dominated by the bulge and/or the bar as shown by \citealt{Pinna2019a, Pinna2019b}, see vertical lines in Fig. \ref{fig:maps_all}) and the outer galaxy (where mono-age populations start flaring). We thus compute these average values of \zhalf by building cumulative vertical mass profiles using all spaxels within $6<|x|<40$~arcsec for FCC~153, $20<|x|<50$~arcsec for FCC~170 and $13<|x|<40$~arcsec for FCC~177.

\section{Results}\label{sec:results} 

\subsection{Global structure of mono-age populations}

\begin{figure*}
\includegraphics[width=\textwidth]{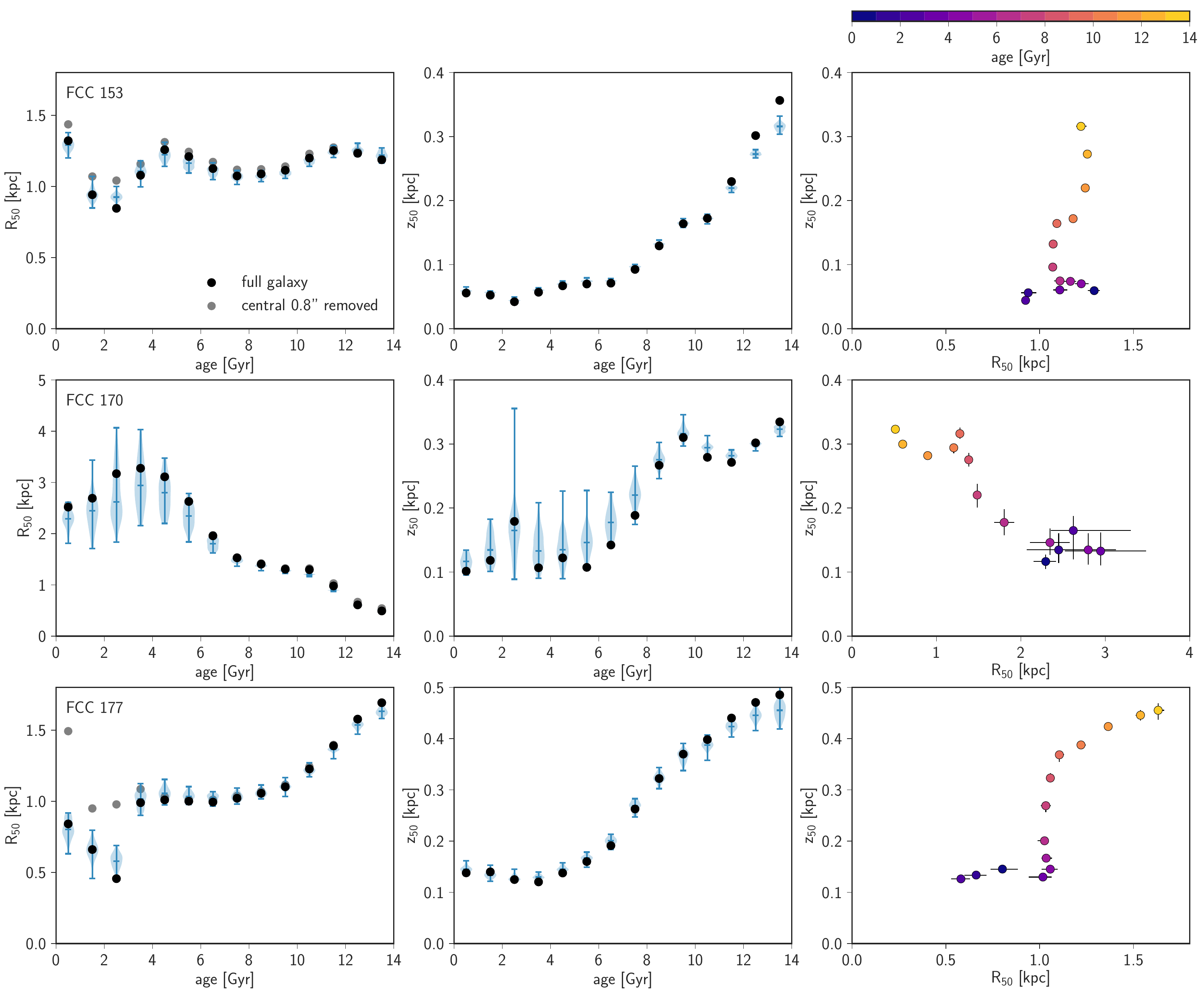}
\caption[]{The first and second columns show \rhalf and \zhalf as a function of age for FCC 153, FCC 170 and FCC 177. The black dots represent the values obtained from our best fit to the spectra, while the blue violin plots represent the distribution of values obtained from the 50 Monte Carlo iterations (the upper and lower limits of the violin plots correspond to the full range of values from the 50 iterations, while the shape of the blue region represents the distribution of the values). The grey dots represent the values from the best fits, but excluding the very central regions of the galaxies, where nuclear star clusters are found. The third column represents \zhalf as a function of \rhalf, color-coded by age, for each galaxy. Here, the error bars represent the 16$^\mathrm{th}$-84$^\mathrm{th}$ percentile range.} 
\label{fig:r50_z50_all}
\end{figure*}

In the first column of Fig. \ref{fig:r50_z50_all}, we show \rhalf as a function of age for mono-age populations in FCC 153, FCC 170 and FCC 177. The black dots represent the values obtained from our best fit to the spectra, while the blue violin plots represent the distribution of values obtained from the 50 Monte Carlo iterations.
For FCC 153, we find a nearly constant \rhalf as a function of age, with only slight variations for stars younger than 3--4 Gyr. We investigate the possible origin of those variations by re-computing \rhalf after excluding the central 0.8 arcsec of the galaxy, where a blue nuclear star cluster (NSC) is found \citep{Turner2012}: these new values are shown with grey dots in Fig. \ref{fig:r50_z50_all}. We find that the NSC cannot fully explain the variations observed in the values of \rhalf for young stars.
In FCC 170, we find that \rhalf increases significantly from old to young populations, with large uncertainties for populations younger than 6 Gyr (this is because of the low mass fractions in those younger populations, see \citealp{Pinna2019a}).
By contrast, in FCC 177, \rhalf is highest for 14 Gyr old stars, then decreases until it reaches a constant value for stars between 8 and 3 Gyr old, and finally drops sharply for younger stars. In this case, the decrease of \rhalf for young stars can be attributed to the presence of a young NSC (also visible in the maps shown in Fig.~\ref{fig:maps_all}): this drop is entirely absent if we exclude the inner 0.8 arcsec. It thus seems that in FCC 177, \rhalf is essentially constant for young and intermediate-age stars within the main body of the galaxy (excluding the NSC). This strong influence of the NSC on the age distribution in the inner region of FCC 177 is consistent with the surface brightness profile shown in \cite{Turner2012}, where the NSC dominates the light in the centre. The three galaxies we study thus show quite different evolutions of \rhalf with age: in FCC 153 \rhalf is nearly constant, in FCC 170 it increases for younger stars, while in FCC 177 it first decreases then stays constant. We note that this might not exactly reflect the behaviour of \rhalf in the overall galaxies as we do not cover the full radial extent of the discs. However, in all cases our observations reach $\sim 70$ per cent of the optical radius, so even though we do not reach the outer disc, we cover the main regions of the galaxies.

While we have just shown that there is some diversity in the radial structures of the galaxies, their global vertical structures are more similar to each other. We show in the middle column of Fig. \ref{fig:r50_z50_all} the evolution of \zhalf as a function of age. In all three cases, \zhalf decreases from old to young stars, with small differences from galaxy to galaxy. In FCC 153, \zhalf first decreases sharply from 14 Gyr old to 8 Gyr old populations, then remains quite constant for younger stars. In FCC 170, the old stars (8--14 Gyr old ) have a nearly constant \zhalf, which then decreases in younger populations (with again large uncertainties for those populations). In FCC 177, \zhalf smoothly decreases from 14 to 6 Gyr old populations, and then remains constant for young stars. We note that the constant \zhalf measured for young stars is not an artificial effect due to a lack of spatial resolution, which would prevent us from detecting thinner discs if they existed. Indeed, in the thinnest galaxy, FCC 153, the young stars have a \zhalf of $\sim 5$ arcsec: this is well above the seeing of our observations, and the vertical extent of the disc is covered by many Voronoi bins (see Fig. \ref{fig:maps_all}).

We summarize our results on the evolution of \rhalf and \zhalf in the right column of Fig. \ref{fig:r50_z50_all}, showing \zhalf as a function of \rhalf color-coded by age. This is reminiscent of the figure produced by \cite{Bovy2012b}, showing the scale-height as a function of scale-length for mono-abundance populations in the Milky Way (their figure 5). In the Milky Way, the scale-height decreases and the scale-length increases when going from old $\alpha$-rich stars to young $\alpha$-poor stars (this is an approximation, because  the relation between mono-abundance and mono-age populations is complex, see \citealp{Minchev2017}). A similar trend is only found here for FCC 170, because, among our galaxies, it is the only one showing an increased \rhalf for younger stars. By contrast, FCC 153 mostly shows a vertical trend in this plane, due to a nearly constant \rhalf as a function of age. Finally, the evolution of FCC 177 seems to correspond to three different phases: first, a decrease of \rhalf with a mild decrease of \zhalf (for stars older than 9 Gyr), then a sharp decrease of \zhalf at nearly constant \rhalf (from 9 to 3 Gyr old), and finally a smaller \rhalf but no change of \zhalf for stars younger than 3 Gyr. This final phase (with a small \rhalf) is connected to the NSC, as previously discussed in this section.

In the next section, we will discuss how \zhalf changes with radius, and how this impacts the global age structure of the three galaxies.

\subsection{Radial structure of mono-age populations and radial age profiles}

\begin{figure*}
\includegraphics[width=\textwidth]{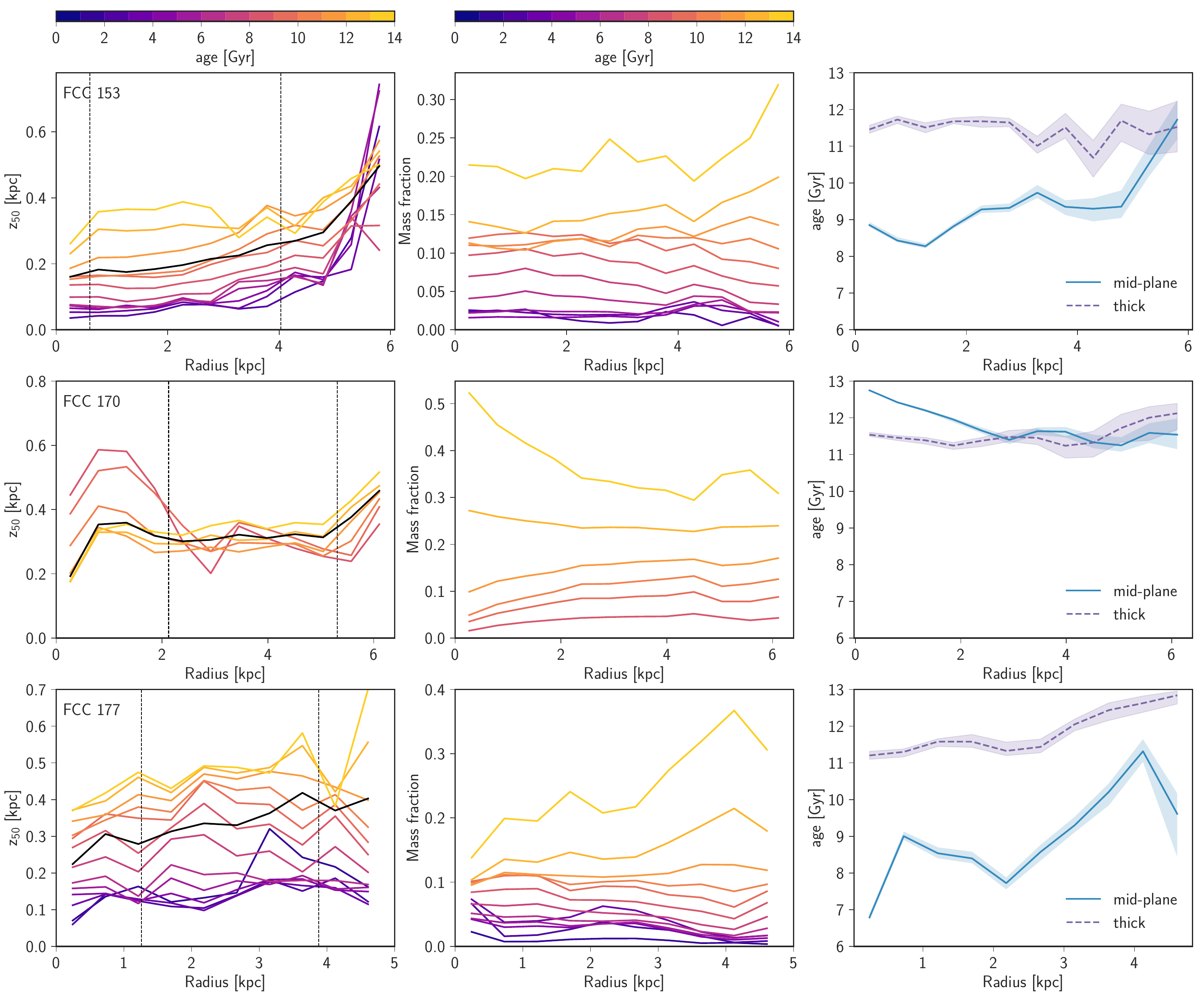}
\caption[]{Structure of FCC 153 (top row), FCC 170 (middle) and FCC 177 (bottom). The first column shows radial profiles of \zhalf for mono-age populations, with each line color-coded by the age of the corresponding population. The thick black line represents the radial evolution of the global \zhalf for each galaxy. The vertical dashed lines delimit the regions of the galaxies we use to compute the mean \zhalf values shown in Fig. \ref{fig:r50_z50_all}. The second column shows the mass fraction in each mono-age population, at each radius. In this column as in the first one, we only plot populations containing at least 1 per cent of the total mass of each galaxy. The last column shows radial profiles of the average mass-weighted age along the mid-plane (blue solid line) and the thick disc (violet dashed line). The shaded area represents the 16-84th percentile range, and the line the median.}
\label{fig:radial_all}
\end{figure*}

We present in the first column of Fig. \ref{fig:radial_all} the radial profiles of \zhalf for mono-age populations in the three galaxies. In each panel, the vertical lines delimit the region we used to compute the average \zhalf values shown in Fig. \ref{fig:r50_z50_all}. In the middle column, we show the fraction of the mass in each population, at each radius. In those first two columns, for clarity, we only represent mono-age populations that contain at least 1 per cent of the total mass of each galaxy (this mostly affects FCC 170, where very few stars are younger than 8 Gyr, see \citealp{Pinna2019a}). Finally, in the third column, we show the average mass-weighted age as a function of radius, computed along the mid-plane (using spaxels within $\pm 1$ arcsec along the major axis), and in thick disc-dominated regions (above 6 arcsec in FCC 153, and 10 arcsec in FCC 170 and in FCC 177, see \citealp{Pinna2019a, Pinna2019b}). 

The upper-left panel of Fig. \ref{fig:radial_all} shows that FCC 153 has a very regular structure: in most regions of the galaxy except the outer disc, \zhalf increases with age, and mono-age populations have mostly flat \zhalf radial profiles, over most of the disc. Flaring starts to be noticed at distances larger than $\sim 4$ kpc (marked by the second vertical line), and flaring is significantly stronger for younger stars: for the youngest stars, \zhalf increases by a factor of $\sim 6$ over 2 kpc, while for the oldest stars \zhalf increases only by a factor of $\sim 1.5$ in that same region. This means that a distance of $5-6$ kpc from the centre of the galaxy, the correlation between \zhalf and age disappears. This strong flaring is also seen in the global \zhalf profile for FCC 153 (black line in the top left panel of Fig. \ref{fig:radial_all}), computed including stars of all ages: this global \zhalf gently increases from the centre to $\sim 5$ kpc, and then sharply increases, reflecting the behaviour of mono-age populations.
The distribution of the mass within each population is also very regular: overall, there is more mass in older populations, at all radii. Outside of 5 kpc, we notice a slight increase in the fraction of very old stars. These mass distributions combine with the radial profiles of \zhalf to produce the average age profiles shown in the upper right panel of Fig. \ref{fig:radial_all}. In the mid-plane, the average age is roughly constant at 8.5--9.5 Gyr  across most of the disc (with a slight positive gradient), and then sharply increases outside of 5 kpc, reaching nearly 12 Gyr at 6 kpc: this increase is due to the combination of a high \zhalf for young stars in this region (decreasing their contribution to the mid-plane), and the higher mass fraction overall of old stars. By contrast, the age profile in the thick disc is essentially flat at 11.5 Gyr.

In FCC 170 (second row of Fig. \ref{fig:radial_all}), the radial profiles of \zhalf first show a bump in the inner region: this follows the shape of the box/peanut bulge in this region. This peak is strongest for 8--9 Gyr old stars. Outside of the inner 2 kpc, populations of all ages have similar \zhalf (note again that we do not show here populations younger than 8 Gyr) and \zhalf  remains essentially flat as a function of radius for all mono-age populations, with a small but coherent increase only in our outermost radial bin. The mass within each population is very regular, with most of the mass in older stars, and we notice an increase in the mass fraction of the very oldest stars in the inner disc. The simplicity of the stellar populations in this galaxy is reflected in the radial age profiles: both the mid-plane and the thick disc have a nearly constant average age of $\sim 11.5$ Gyr. There is a small increase of the mean age towards the centre of the galaxy, in the mid-plane: this corresponds to the smaller \zhalf and larger mass fraction for the oldest stars within the inner 2 kpc of the galaxy.

Finally, in FCC 177 (bottom row of Fig. \ref{fig:radial_all}), we also find very flat  and regular radial profiles for the \zhalf of mono-age populations, with no sign of flaring. On the other hand, the global \zhalf increases very slightly as a function of radius: this is due to the increased dominance of old stars (with a larger \zhalf) in the outer disc. This can be seen in the middle panel of the last row in Fig. \ref{fig:radial_all}: stars older than 12 Gyr represent $\sim 25$  per cent of the mass in the centre of the galaxy, and 50  per cent in the outer disc. This means that we see radial age gradients in FCC 177, both in the mid-plane and in the thick disc.

By comparing those three galaxies, we confirm that the overall structure of the disc (in terms of global \zhalf and age gradients) is determined both by the radial profiles of \zhalf for mono-age populations, and by the mass contributed by each mono-age population at each radius.

\subsection{Spatial distribution of metal-rich and metal-poor stars} \label{subsec:spat_distr}
\begin{figure*}
\includegraphics[width=\textwidth]{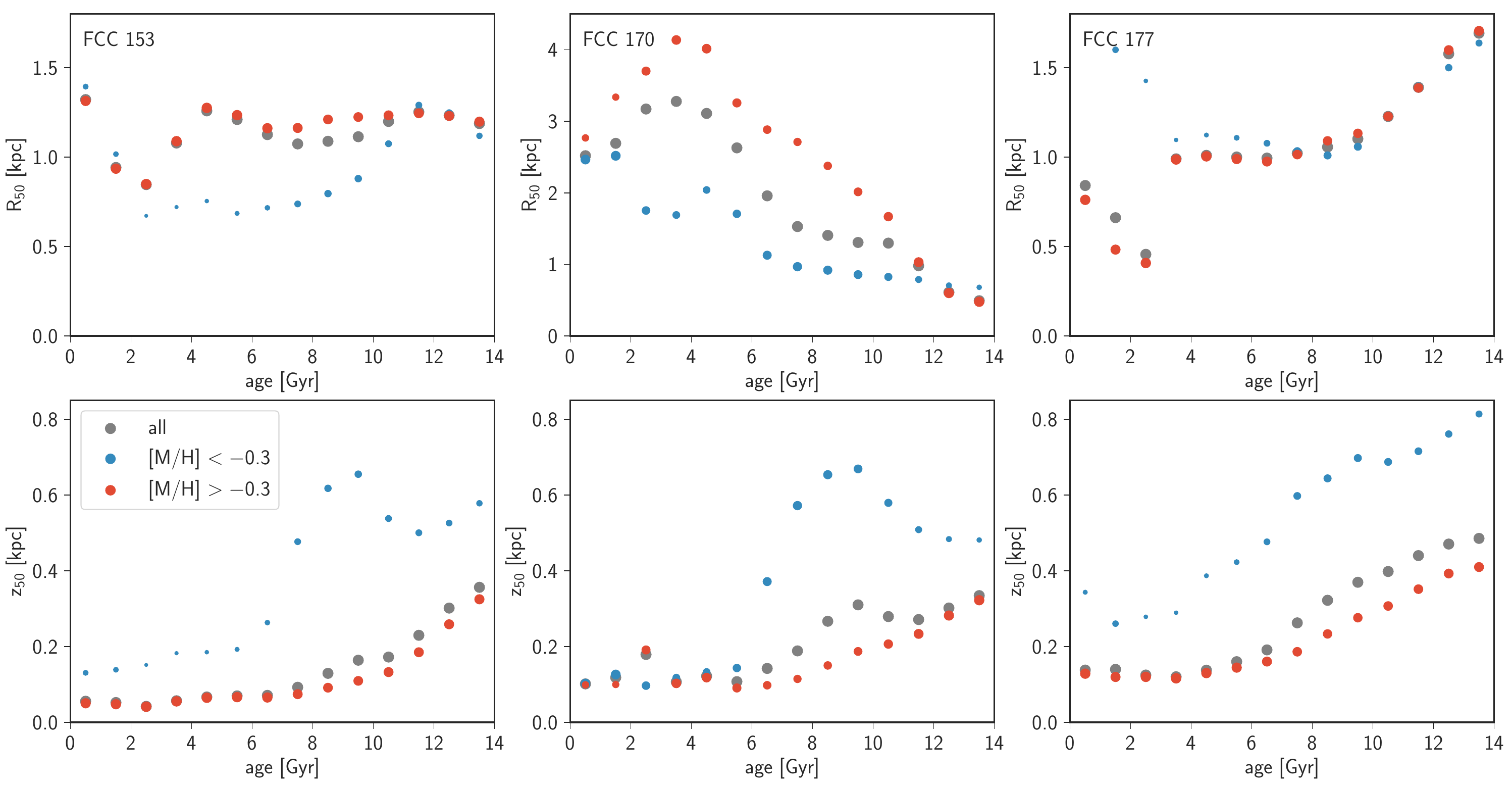}
\caption[]{\rhalf (top row) and \zhalf (bottom row) as a function of age for FCC 153, FCC 170 and FCC 177. The grey dots represent the values obtained for the global population (they are the same as the black dots in Fig. \ref{fig:r50_z50_all}), while the blue and red dots correspond to metal-poor and metal-rich stars, respectively. The surface of each blue and red dot is proportional to the fraction of the mass in this metallicity range at a given age. \cite{Pinna2019a,Pinna2019b} proposed that the 8--11 Gyr old metal-poor stars could be accreted from small galaxies: we indeed find that these stars have a different spatial distribution compared to the metal-rich stars (possibly formed in-situ).} 
\label{fig:z50_r50_metal_split}
\end{figure*}
\cite{Pinna2019a,Pinna2019b} discovered complex stellar populations in these three galaxies. In particular, they found that in the thick discs, 8--11~Gyr old stars are on average more metal-poor than older stars. One possible interpretation is to identify those younger metal-poor stars as accreted from a low mass galaxy. Indeed, lower mass galaxies are overall more metal-poor, so at a given age we would expect accreted stars to be more metal-poor than stars born in-situ (see also \citealp{Boecker2020, Martig2021, Davison2021} for similar ideas). \cite{Pinna2019a,Pinna2019b} find that in all three galaxies those accreted stars would have mean ages between
8 and 11 Gyr, a mean metallicity of -0.6 and a mean [Mg/Fe] of around 0.35, and would represent 4--5  per cent of the total mass of the galaxies.

To uncover the possible effects of those past mergers on the structure of the three galaxies, we re-compute \rhalf and \zhalf for metal-rich and metal-poor stars separately. We place a somewhat arbitrary limit between the two components at [M/H]$=-0.3$, as a rough separation between metal-poor accreted stars and metal-rich in-situ stars (the reality is surely more complex, we only aim here to see a difference between two populations of stars). We show in Figure \ref{fig:z50_r50_metal_split} the evolution of \rhalf and \zhalf as a function of age for the overall population, as well as for the metal-poor and metal-rich stars. As expected, most of the overall trends observed are driven by the trends seen in the metal-rich (possibly in-situ) stars, which are the most abundant in those galaxies. However, it is also interesting to notice that, at a given age, metal-rich and metal-poor stars tend to have different spatial distributions. In all cases, metal-poor stars are thicker than metal-rich stars, and they also tend to be more centrally concentrated (except in FCC 177 where all stars have the same radial distribution, even though they have different vertical distributions). We also find that, while the metal-rich stars show a very smooth and regular  increase of \zhalf with age, the metal-poor stars have a more complex evolution, with an increased \zhalf for populations 8--10 Gyr old. This corresponds to the time of the mergers initially proposed by \cite{Pinna2019a, Pinna2019b}.

We thus suggest that stars from those accreted galaxies were deposited in quite thick and disc-like configurations (with a smaller radial extent than in-situ stars, except in FCC 177), and also that the in-situ stars have not been significantly perturbed by the interaction: their \zhalf profiles rise very smoothly with age and show no sign of  a sharp increase at the time of the interaction.

\section{Discussion}\label{sec:discussion}

\subsection{Stellar population analysis}

The same MUSE data was independently analysed by \cite{Poci2021} and \cite{MartinNavarro2021}, using different setups of pPXF (e.g., different order of the multiplicative polynomial and regularization parameter used for the fit), different SSP libraries, and different wavelength ranges for the fits. In spite of this, the resulting age and metallicity maps are very similar between all studies: the main difference is that \cite{Poci2021} and \cite{MartinNavarro2021} show light-weighted quantities (while we use mass-weighted quantities), so that, as expected, their mean ages are 1--2 Gyr younger than ours. \cite{Ding2023} also verified that the age and metallicity gradients are similar in the maps presented by \cite{MartinNavarro2021} and \cite{Pinna2019a,Pinna2019b}, used here.

While the recovery of the average ages in each Voronoi bin (shown in the maps), is relatively stable, recovering full star formation histories is a much more complex endeavour: this is an ill-conditioned problem because of strong degeneracies between spectra of different stellar populations \citep[e.g.,][]{Ocvirk2006,Zibetti2024}. After analysing in detail the information content within spectra, \cite{Zibetti2024} recommend that the age bins used to extract star formation histories should have a relative width of at least 30  per cent: this means a bin width of $\sim$ 3 Gyr at old ages. For our mono-age populations, we chose a width of only 1 Gyr, which might seem too narrow, particularly at old ages where spectra vary very little with age. However, our analysis shows that some age information is present in the spectra, even at old ages: we find a smooth increase of \zhalf as a function of age (particularly for FCC 153 and FCC 177), even for populations older than 9 Gyr. Note also that we fit every spatial bin independently, but still recover smooth variations in \zhalf across the galaxies (Fig. \ref{fig:radial_all}). This justifies a posteriori the validity of our choice of 1 Gyr-wide age bins.

Still, we acknowledge that the uncertainties we report for \zhalf and  \rhalf in Fig. \ref{fig:r50_z50_all} are probably severely underestimated, as they only correspond to statistical uncertainties. Systematic uncertainties due to details of the fitting technique (e.g., order of the multiplicative polynomial, or regularization parameter) have been shown by \cite{Pinna2019a} to be relatively small, but the properties of the SSPs used have a much stronger impact on the results. Exploring this fully is a difficult and costly task, far beyond the scope of this paper. We thus caution the reader against putting too much trust into the absolute values we measure for \rhalf and \zhalf, but point out that the general trends we derive are more robust.

\subsection{Projection effects}

The measurement of the structure of edge-on galaxies is affected by projection effects, particularly towards the centre of galaxies where in the same line-of-sight, stars from the outer disc are superimposed on stars from the inner disc. This might have an impact on the measurement of a global \rhalf, as well as on any measurement of radial gradients.
\cite{Ibarra2019} explored this effect by using simulated galaxies and showed that age profiles are artificially flattened in highly inclined galaxies. We also expect that the same geometric effect could in some cases artificially flatten the radial profiles we measure for \zhalf. Indeed, if a disc is intrinsically flared, towards the centre of the galaxy we observe thick outer disc populations superimposed on thin inner disc populations: in such a case, the thickness we measure in the inner region is overestimated.

It is possible to estimate the full 3D structure of galaxies by using dynamical models, as done for our three galaxies by \cite{Poci2021} and \cite{Ding2023}. While powerful, these dynamical models are also associated of course with additional uncertainties. Still we can compare our projected radial age gradients to the intrinsic age gradients shown by \cite{Ding2023} using the full orbital structure of the galaxies.
For FCC 153, they find a mostly flat overall intrinsic radial age profile, with a decrease in the centre of the galaxy. This decrease is present both in the profiles for stars on cold and hot orbits, but is strongest for the cold disc, where a clear age decrease is seen in the inner $\sim 1.5$ kpc. This matches the younger ages we see in projection for stars in the mid-plane in the inner $\sim 1.5$ kpc.
For FCC 170, \cite{Ding2023} report uniformly old ages throughout the galaxy and flat age profiles for cold and hot stars, with possibly a very small age increase towards the centre for stars on hot orbits. This also corresponds well to what we see in the projected profiles.
Finally, for FCC 177 they find a mostly flat intrinsic age profile overall, with a clear decrease at radii smaller than 1 kpc. Stars on cold orbits have a flat age profile, while stars on hot orbits show a positive age gradient throughout the disc, with a strong drop of ages in the inner 1 kpc. They note that this is partially due to the young nuclear star cluster, which contributes to the hot component. We also see this positive age gradient in our projected data.

We can also compare the vertical structure we infer for mono-age populations to the measurements presented by \cite{Poci2021}. They determine the vertical velocity dispersion ($\sigma_z$) as a function of age and metallicity, and find a systematic increase of $\sigma_z$ with age, which is consistent with the increase of \zhalf with age we show in Fig. \ref{fig:r50_z50_all}. They also find that at fixed age, metal-poor stars have a larger $\sigma_z$, which matches the results we show in Fig. \ref{fig:z50_r50_metal_split} of an increased \zhalf for metal-poor stars.

This means that while projection effects might flatten the gradients we measure, the general agreement between our results and those coming from dynamical models is reassuring.

\subsection{Vertical heating of the disc}
While the number of existing observational studies on disc thickness as a function of age is limited, it is useful to compare our results to what has been found for the Milky Way and nearby galaxies.

First, we find that the $\sim$1 Gyr old stars have a \zhalf between 50 and 150 pc (Fig. \ref{fig:r50_z50_all}, middle column). While other studies often report values of scale-heights obtained from fitting exponential density  profiles  (where the scale-height is called $h_z$) or fitting sech$^2$ profiles (where the scale-height is called $z_0$), we can convert those values into half-mass heights: for an exponential density profile, $z_{\mathrm{50}}= \ln(2)h_z$, and $h_z=z_0/2$ \citep{vanderKruit1981}. Using these conversions, in the Milky Way, 2 Gyr old stars have a \zhalf of $\sim 150$ pc \citep{Mackereth2017}, while in the three edge-on galaxies studied by \cite{Streich2016} the youngest stars (with ages typically below 0.5 Gyr) have a \zhalf between 70 and 140 pc. The values we find are thus quite typical of young stars in the Milky Way and other nearby galaxies.

The general increase of thickness with age is also consistent with observations of other nearby galaxies \citep{Seth2005, Tikhonov2005, Dorman2015, Streich2016} and of the Milky Way \citep{Mackereth2017, Xiang2018}. In the Milky Way, the most studied relation is actually the correlation between vertical velocity dispersion and age: the vertical velocity dispersion rises smoothly as a function of age, following a power law form $\sigma_z \propto t^\beta$ with $\beta$ between 0.4 and 0.5 \citep{Nordstrom2004,Aumer2009,Sanders2018, Anders2023}. In addition to this smooth increase, there is possibly a jump of $\sigma_z$ to larger values at old ages \citep{Quillen2001,Anders2023}.
 We can further compare our results to these observations by noting that, at vertical equilibrium, the vertical velocity dispersion for stars distributed in an isothermal sech$^2$ density profile with a scale-height $z_0$ is 
$$\sigma_z^2=2 \pi G \Sigma z_0$$ where $\Sigma$ is the surface density \citep{vanderKruit1988}. This means that if $\sigma_z \propto t^\beta$, then we would expect $z_0 \propto t^{2 \beta}$. In summary, if vertical heating happened similarly in our galaxies as in the Milky Way, we would observe \zhalf to increase as a function of age as a power law with an exponent between 0.8 and 1.

What we find is very different: for none of our three galaxies the relation between  \zhalf and age is a single power law. Instead, in FCC 153, FCC 170 and FCC 177 \zhalf is relatively constant for stars from 0 to $\sim 6$ Gyr old, followed by an increase towards older ages. The constant \zhalf for an extended period of time is consistent with an absence of secular disc heating in these galaxies. Indeed, interactions of disc stars with Giant Molecular Clouds \citep{Spitzer1951, Spitzer1953} or with the spiral structure in a galaxy \citep{Carlberg1985} are possible sources of secular heating, but our lenticular galaxies do not contain massive molecular clouds (they are undetected in HI and CO, see \citealp{Zabel2019} and \citealp{Loni2021}), and probably have a very weak spiral structure. Given their low star formation rates for the past few Gyrs, it is very likely that their molecular gas content has been low for a while, with few massive clouds. This means that sources of secular heating are limited, and explains why \zhalf can remain constant over several Gyr. This result is still somewhat surprising given the dense environment our galaxies live in. In particular, simulations show that when galaxies orbit within a cluster, tidal shocking at pericentre passage thickens (and possibly destroys) cold discs \citep{Joshi2020, Galan2022, Ding2024}. Galaxy harassment is also expected to drive morphological changes in disc galaxies \citep{Moore1998}. Our three galaxies have likely resided within the Fornax cluster for longer than 6 Gyr \citep{Iodice2019a,Ding2023}: the constant \zhalf we observe for stars over this long period of time suggests that tidal shocking and harassment have not been efficient at heating the cold discs, at least within their inner regions.

On the other hand, we see an increase of \zhalf for older stars: some of this heating could come from  tidal shocking and harassment. However, 10 Gyr ago, the Fornax cluster might have only assembled 10 per cent of its final mass \citep{Chiang2013}, so that any environmental effects would have been weaker at that time (and, at least for FCC 177, the infall into Fornax is more recent than this, see \citealp{Ding2023}). In addition to this, we measure  a different thickness for metal-poor and metal-rich stars in this age range (see Fig. \ref{fig:z50_r50_metal_split}): there is no reason why  tidal shocking and harassment should affect differently stars of different metallicities. On the other hand, as discussed in Section~\ref{subsec:spat_distr}, this period of increased heating coincides with possible mergers happening 8--10 Gyr ago in those galaxies. One interpretation is that part of the increase in thickness comes from the presence of metal-poor accreted stars, being deposited throughout the galaxies in a thick configuration, and the rest comes from heating of the in-situ metal-rich stars. We do not observe a sharp jump in \zhalf, corresponding to the heating of pre-existing disc stars at the time of the merger, but a jump could be erased by age uncertainties \citep{Martig2014b}, so we cannot draw strong conclusions from the absence of such a jump. Alternatively, the high \zhalf for old in-situ stars and its gradual decrease could also be due to disc settling, as galaxies transition from an early phase with a high star formation rate and a more turbulent disc, to a quiescent phase where stars form in thinner configurations  \citep{Bird2013,Kassin2014, Grand2016, Ma2017, Ceverino2017, Dubois2021, vanDonkelaar2022}.

Finally, another possible difference between our measurements and other observations for nearby galaxies is the magnitude of the increase in thickness from young to old stars. Indeed, in the galaxies studied by \cite{Streich2016}, old stars are 1.5 to 3 times thicker than young stars, while in the Milky Way, old stars are 3 times thicker than young stars \citep{Mackereth2017}. In our three galaxies, we find that old stars are between 3 and 5 times thicker than young stars, which would be consistent with a more violent history in earlier type galaxies. We note that \cite{Yoachim2006} report that  thick discs are 2 to 5 times thicker than thin discs in early type galaxies: this would be consistent with our results if we associate old stars with the thick disc and young stars with the thin disc. The large difference in thickness we observe between young and old stars suggests that mergers, rather than simply disc settling, might be the reason for the increased thickness of older stars.

\subsection{Strength and origin of disc flaring}

The overall amount of disc flaring varies between the three galaxies (see black lines in the left column of Fig. \ref{fig:radial_all}). In FCC 153, the global \zhalf first increases slowly (from a value of 0.18 kpc in the inner disc to 0.25 kpc at a distance of 4 kpc from the centre), then doubles over the final 2 kpc. In FCC 170, the global \zhalf remains constant at 0.3 kpc for most of the disc, and rises by 50 per cent  over the final kpc we observe. In FCC 177, the increase of \zhalf is more gradual, going from 0.3 kpc in the inner disc, to 0.4 kpc in the outskirts.
This amount of flaring that we measure is possibly decreased by projection effects, as discussed in the previous section, but also by the limited radial extent of the MUSE data, which only reaches about 70 per cent of the $B$-band optical radius of those three galaxies \citep[RC3,][]{deVaucouleurs1991}: this means that we might be missing flaring happening in the very outer disc. We can estimate what we might be missing by comparing our results to the analysis done by \cite{Iodice2019b}, using  deep $r$-band images from the Fornax Deep Survey. After subtracting from those images a two-dimensional fit of the isophotes, they study the shape of the residuals: they detect some flaring in all three galaxies. In the $r$-band, the amount of disc flaring is similar in FCC 153 and FCC 170, and much weaker in FCC 177. This matches well what we deduce from our analysis of the global \zhalf in those three galaxies, and suggests that the very mild level of flaring we observe in FCC 177 is not due to the limited radial extent of the MUSE data.

The flaring we observed in the global \zhalf results from the complex combination of the \zhalf of individual mono-age populations, and the contribution from each population at each radius (as discussed for simulated galaxies by \citealp{Minchev2015} and \citealp{Garcia2021}). FCC 170 has the simplest structure: all populations shown in Fig. \ref{fig:radial_all} have a very similar structure, which is then followed by the global \zhalf. This results in a uniform age structure for the galaxy, with no age gradient in the thick disc.
In FCC 153, younger stars flare more than older stars, but their mass fraction is low in the outer disc. This means that these younger stars do not contribute significantly to the thick disc, so that the global   \zhalf follows the shape of the older populations and no age gradient is present in the thick disc.
Finally, in FCC 177, the individual mono-age populations have flatter radial profiles of  \zhalf than the global disc. A global flaring is seen because old stars, which have a larger \zhalf, significantly increase their contribution in the outer disc. This results in positive age gradients, both in the mid-plane and the thick disc. None of these situations matches exactly a simulated galaxy presented in \cite{Minchev2015}, \cite{Garcia2021} or \cite{Sotillo2023}. In particular, all the simulated  galaxies in \cite{Garcia2021} have negative age gradients in their mid-plane, which is very different from what we observe. However, these simulations are focused on Milky Way- or M31-mass spiral galaxies outside of dense environments: it is not surprising that lenticulars in the Fornax cluster would have a different structure. 

Still, the diversity of flaring structure we see in our three galaxies is reminiscent of the diversity discussed in \cite{Sotillo2023}: for simulated galaxies in TNG50, there are cases where young populations are the most flared, cases where it is the opposite, and cases with very little flaring. Observational data on this topic is extremely rare: for three low mass galaxies, \cite{Streich2016} show that the youngest stars have a constant scale-height as a function of radius, while older populations exhibit a mild flaring. For the Milky Way, while several papers have shown that mono-age populations flare, there are disagreements on the amount of flaring in populations of various ages  (see \citealp{Bovy2016, Mackereth2017,Imig2025, Khoperskov2025}). 

Unveiling the cause of the flaring we observe is not trivial. Mergers can cause discs to flare \citep{Bournaud2009}, but this effect can be suppressed if mergers are accompanied by large-scale radial migration \citep{Minchev2012}, so that there is no one-to-one correlation between merger activity and amount of flaring \citep{Garcia2021, Sotillo2023}. While all our galaxies are expected to have undergone mergers 8--10 Gyr ago, their old populations are not strongly flared, which could be consistent with radial migration suppressing flaring during mergers \citep{Minchev2012}. For stars younger than 8 Gyr, we have no evidence of mergers affecting the galaxies, but we still find some flaring, particularly in the very outer regions of FCC~153. One possibility that would fit with our observation that when mono-age populations flare, they only flare in their very outer parts, is that the flaring is due to interactions with dark matter substructures \citep{Yurin2015}: this can heat discs in their outer parts, while leaving the inner regions intact. The different amount of flaring seen in the three galaxies would then correspond to different amounts of interactions, due to different positions of the three galaxies in the cluster.

\subsection{Disc growth and star formation quenching}

All three galaxies have star formation histories peaking at high redshift, with some residual star formation down to the present time. FCC 170 is the most extreme case: most of its stars are older than 8 Gyr \citep{Pinna2019a, Pinna2019b}.   This is consistent with the different environments the galaxies live in: FCC 170 is in a high-density region of the Fornax cluster (the north-south clump), FCC 177 is further away from the core but still in the region traced by X-ray emission, and FCC 153 (the bluest galaxy) is furthest from the centre, although still within the virial radius of Fornax \citep{Iodice2019a}. 

The low levels of star formation at late times are consistent with the very low gas fractions found in the galaxies: none of the galaxies is detected in CO \citep{Zabel2019} or HI, down to a 3 sigma sensitivity of $\mathrm{M_{HI}}= 5\times 10^5\ \mathrm{M_\odot}$ \citep{Loni2021, Kleiner2025}. This low gas content could be due to ram-pressure stripping, removing the existing cold gas from the galaxies as they move through the hot intracluster medium \citep{Gunn1972}. Another possibility is strangulation, where galaxies simply consume their existing gas, while being shut off from their supply of cold gas \citep{Larson1980}.
Fornax is a relatively low mass cluster, so that ram-pressure stripping is expected to be less efficient than in a cluster like Virgo. However, 
using Fornax-analogues in the TNG50 simulation, \cite{Ding2024} show that within 4 Gyr of galaxies falling into a cluster, 2/3 of galaxies see the radius of their star-forming gas disc decreasing by half. This is enough to sharply decrease star formation in the outer disc of simulated galaxies. Using cold molecular gas observations of galaxies within Fornax, \cite{Zabel2019} have indeed shown that some galaxies are currently being affected by ram-pressure stripping.
 
However, for the three galaxies we study here, our measurements of the radial structure of mono-age populations do not show strong indications of ram-pressure stripping within the radial range we probe. In particular, in FCC 170 (which is in the densest region of Fornax, so is possibly the most affected by its environment), \rhalf increases for younger stars, which is the opposite of what is expected if quenching were due to ram pressure stripping. For FCC 177, we see a decrease of \rhalf at very early times (for ages above 10 Gyr). This early decrease could be consistent with ram-pressure stripping, but \cite{Ding2023} predict an infall time of FCC 177 into the cluster $6\pm2$ Gyr ago. If true, this means that ram-pressure stripping cannot explain the size evolution of stellar populations born 10-13 Gyr ago. In any case, this probably corresponds to a time when the cluster would have been much less massive, so it is unclear if ram-pressure stripping could anyway be responsible for the early evolution of \rhalf in this galaxy. On the other hand, for the past 8 Gyr, we find a nearly constant \rhalf for mono-age populations in both FCC 177 and FCC 153: this is not consistent with a gradual truncation of the star-forming gas disc because of ram-pressure stripping (it is still possible that stripping affected the very outer regions of the galaxies).

Even if it is not due to ram-pressure stripping, this nearly constant \rhalf as a function of age is still probably due to an effect of the environment within Fornax. Indeed, normal inside-out formation would predict younger stars to be more extended and have larger scale-lengths than older stars \citep{Bird2013, Stinson2013, Martig2014a, Buck2020}, and this is what is observed for most galaxies \citep{Williams2009, Gogarten2010,Peterken2020} and for the Milky Way \citep{Amores2017, Xiang2018}. A constant scale-length as a function of age could be due to an extended series of mergers \citep{Buck2020} but there are no signs of recent mergers in the stellar populations of the galaxies \citep{Pinna2019a, Pinna2019b}, and no signs of accretion or disruption in deep images \citep{Iodice2019b}. Alternatively, very strong radial migration could cause mono-age populations to end up with similar scale-lengths \citep{Frankel2019}, but radial migration is probably weak in these lenticular galaxies with thick discs and probably weak spiral structures.

Our favourite interpretation for the recent suppression of star formation, and the constant \rhalf of mono-age populations as a function of age is simply strangulation: the slow consumption of the existing gas, in the absence of a  supply of fresh cold gas. This means that the radial extent of the star-forming gas disc remains constant as a function of time, and that mono-age populations have then a very similar radial extent over many Gyrs. We simply  observe the absence of inside-out growth, because there is no supply of fresh gas with a high angular momentum.

As a conclusion, our observations cannot rule out an effect of ram-pressure stripping on the outer disc of FCC 153, FCC 170, and FCC 177, but the structure of the inner regions is most likely due to strangulation slowly decreasing the overall gas content while keeping constant  the radial extent of the star-forming gas.

 \section{Conclusion}

Using deep MUSE observations of three edge-on lenticular galaxies in the Fornax cluster, we study the spatial distribution of stars in mono-age populations. In particular, we measure the half-mass radius (\rhalf) as well as the half-mass height (\zhalf) for populations within 1~Gyr wide age bins. In all three galaxies, we find that \zhalf is relatively constant for stars from 0 to $\sim$6 Gyr old, followed by an increase towards older
ages. The radial profiles of \zhalf  for individual mono-age populations are mostly flat (or mildly flared) in all galaxies, with an increased flaring in the outer disc for FCC 153 and FCC 170. On the other hand, the galaxies have a different radial structure: FCC 153 has a constant \rhalf as a function of age, FCC 170 shows an increasing \rhalf with decreasing age, while in FCC 177, \rhalf is highest for old stars, then decreases before staying constant for stars younger than 8 Gyr.

Based on a detailed analysis of stellar populations in these galaxies, \cite{Pinna2019a,Pinna2019b} proposed that all three galaxies experienced some mergers 8--10 Gyr ago: we find further hints supporting this scenario in the different spatial distribution of metal-poor (possibly accreted) and metal-rich (possibly in-situ) stars. We find that accreted stars were deposited in quite thick and disc-like configurations, with a smaller radial extent than in-situ stars, except in FCC 177. We also find that those mergers had a limited impact on the structure of pre-existing stars: it seems that the in-situ stars have not been significantly perturbed by the interaction, since their \zhalf profiles rise very smoothly with age. We also do not see strong flaring in old populations present in the galaxy at the time of the mergers.

Apart from this increase of \zhalf at old ages, we find that the thickness of mono-age populations has remained remarkably constant in all three galaxies for the past $\sim 6$ Gyr. This is consistent with an absence of secular disc heating in these galaxies, possibly due to the near absence of heating agents such as Giant Molecular Clouds or spiral arms. This also means that external sources of heating such as tidal shocking (when galaxies are at the pericentre of their orbit within the cluster) or galaxy harassment have been inefficient in these galaxies, allowing them to keep thin discs long after their infall into the cluster.

The effect of the cluster environment is thus not obvious on the vertical structure of the galaxies, but it is clear in the global star formation histories, and the radial distribution of stars.
All three galaxies have star formation histories peaking at high redshift, with some residual, low-level star formation down to the present time. We cannot rule out an effect of ram-pressure stripping on the very outer regions of the galaxies, outside of our field of view. However, within the regions we observe, ram-pressure stripping is probably not the main source of gas removal and quenching, as we find that mono-age populations have a radial extent that remained constant for the past 8 Gyr. Our favourite interpretation for the recent suppression of star formation, and the constant \rhalf of mono-age populations as a function of age is simply strangulation: the slow consumption of the existing gas, in the absence of a  supply of fresh cold gas. This means that the radial extent of the star-forming gas disc remains constant as a function of time, and that mono-age populations have then a very similar radial extent over many Gyrs. We simply observe the absence of inside-out growth, because there is no supply of fresh gas with a high angular momentum. The effect of the cluster environment on our three galaxies is thus rather indirect: by limiting the supply in fresh gas, it suppresses galaxy growth, and also indirectly suppresses secular heating by limiting the amount of heating agents within the discs.

On a final note, we would like to highlight the usefulness of using IFS data and full spectral fitting to study the detailed structure of mono-age populations. This gives us rich insights on how the global structure of the galactic discs results from the complex superposition of individual mono-age populations. In particular, we find that global radial age gradients and the global shape of the disc are not necessarily good tracers of the underlying structure of stars. This is because these global observables are determined both by the radial profiles of \zhalf for mono-age populations, and by the mass contributed by each mono-age population at each radius. In future studies, we plan to extend our analysis to a larger sample of galaxies from the GECKOS survey \citep{vandeSande2024}. With deep MUSE observations of 36 nearby, edge-on, Milky Way-mass galaxies, GECKOS is the perfect sample to reveal the full diversity of galaxy structures along the Hubble sequence, and should give us rich insights into the build-up of disc galaxies.

\section*{Acknowledgements}
We are grateful to the Fornax3D team for the excellent quality of the data they are publicly sharing.
This work is based on observations collected at the European Organization for Astronomical Research in the Southern Hemisphere under ESO program 296.B-5054(A).
MM and YD acknowledge support from the UK Science and Technology Facilities Council through grant ST/Y002490/1.
FP acknowledges support from the Horizon Europe research and innovation programme under the Maria Skłodowska-Curie grant “TraNSLate” No 101108180, and from the Agencia Estatal de Investigación del Ministerio de Ciencia e Innovación (MCIN/AEI/10.13039/501100011033) under grant (PID2021-128131NB-I00) and the European Regional Development Fund (ERDF) ``A way of making Europe''. 
FP wishes to acknowledge the contribution of the IAC High-Performance
Computing support team and hardware facilities to the results of this research. IM acknowledges support by the Deutsche Forschungsgemeinschaft under the grant MI 2009/2-1.
 
\section*{Data Availability}

The Fornax3D MUSE data cubes are accessible via the ESO archive. All the information can be found \href{https://www.eso.org/sci/publications/announcements/sciann17540.html}{here}.



\bibliographystyle{mnras}
\bibliography{library_F3D} 

\begin{thebibliography}{}
\makeatletter
\relax
\def\mn@urlcharsother{\let\do\@makeother \do\$\do\&\do\#\do\^\do\_\do\%\do\~}
\def\mn@doi{\begingroup\mn@urlcharsother \@ifnextchar [ {\mn@doi@} {\mn@doi@[]}}
\def\mn@doi@[#1]#2{\def\@tempa{#1}\ifx\@tempa\@empty \href {http://dx.doi.org/#2} {doi:#2}\else \href {http://dx.doi.org/#2} {#1}\fi \endgroup}
\def\mn@eprint#1#2{\mn@eprint@#1:#2::\@nil}
\def\mn@eprint@arXiv#1{\href {http://arxiv.org/abs/#1} {{\tt arXiv:#1}}}
\def\mn@eprint@dblp#1{\href {http://dblp.uni-trier.de/rec/bibtex/#1.xml} {dblp:#1}}
\def\mn@eprint@#1:#2:#3:#4\@nil{\def\@tempa {#1}\def\@tempb {#2}\def\@tempc {#3}\ifx \@tempc \@empty \let \@tempc \@tempb \let \@tempb \@tempa \fi \ifx \@tempb \@empty \def\@tempb {arXiv}\fi \@ifundefined {mn@eprint@\@tempb}{\@tempb:\@tempc}{\expandafter \expandafter \csname mn@eprint@\@tempb\endcsname \expandafter{\@tempc}}}

\bibitem[\protect\citeauthoryear{{Agertz} et~al.,}{{Agertz} et~al.}{2021}]{Agertz2021}
{Agertz} O.,  et~al., 2021, \mn@doi [\mnras] {10.1093/mnras/stab322}, \href {https://ui.adsabs.harvard.edu/abs/2021MNRAS.503.5826A} {503, 5826}

\bibitem[\protect\citeauthoryear{{Am{\^o}res}, {Robin}  \& {Reyl{\'e}}}{{Am{\^o}res} et~al.}{2017}]{Amores2017}
{Am{\^o}res} E.~B.,  {Robin} A.~C.,   {Reyl{\'e}} C.,  2017, \mn@doi [\aap] {10.1051/0004-6361/201628461}, \href {https://ui.adsabs.harvard.edu/abs/2017A&A...602A..67A} {602, A67}

\bibitem[\protect\citeauthoryear{{Anders} et~al.,}{{Anders} et~al.}{2023}]{Anders2023}
{Anders} F.,  et~al., 2023, \mn@doi [\aap] {10.1051/0004-6361/202346666}, \href {https://ui.adsabs.harvard.edu/abs/2023A&A...678A.158A} {678, A158}

\bibitem[\protect\citeauthoryear{{Aumer} \& {Binney}}{{Aumer} \& {Binney}}{2009}]{Aumer2009}
{Aumer} M.,  {Binney} J.~J.,  2009, \mn@doi [\mnras] {10.1111/j.1365-2966.2009.15053.x}, \href {https://ui.adsabs.harvard.edu/abs/2009MNRAS.397.1286A} {397, 1286}

\bibitem[\protect\citeauthoryear{{Bacon} et~al.,}{{Bacon} et~al.}{2010}]{Bacon2010}
{Bacon} R.,  et~al., 2010, in {McLean} I.~S.,  {Ramsay} S.~K.,   {Takami} H.,  eds,  Society of Photo-Optical Instrumentation Engineers (SPIE) Conference Series Vol. 7735, Ground-based and Airborne Instrumentation for Astronomy III. p. 773508, \mn@doi{10.1117/12.856027}

\bibitem[\protect\citeauthoryear{{Beasley}, {San Roman}, {Gallart}, {Sarajedini}  \& {Aparicio}}{{Beasley} et~al.}{2015}]{Beasley2015}
{Beasley} M.~A.,  {San Roman} I.,  {Gallart} C.,  {Sarajedini} A.,   {Aparicio} A.,  2015, \mn@doi [\mnras] {10.1093/mnras/stv943}, \href {https://ui.adsabs.harvard.edu/abs/2015MNRAS.451.3400B} {451, 3400}

\bibitem[\protect\citeauthoryear{{Bedregal}, {Arag{\'o}n-Salamanca}, {Merrifield}  \& {Milvang-Jensen}}{{Bedregal} et~al.}{2006}]{Bedregal2006}
{Bedregal} A.~G.,  {Arag{\'o}n-Salamanca} A.,  {Merrifield} M.~R.,   {Milvang-Jensen} B.,  2006, \mn@doi [\mnras] {10.1111/j.1365-2966.2006.10829.x}, \href {https://ui.adsabs.harvard.edu/abs/2006MNRAS.371.1912B} {371, 1912}

\bibitem[\protect\citeauthoryear{{Bedregal}, {Cardiel}, {Arag{\'o}n-Salamanca}  \& {Merrifield}}{{Bedregal} et~al.}{2011}]{Bedregal2011}
{Bedregal} A.~G.,  {Cardiel} N.,  {Arag{\'o}n-Salamanca} A.,   {Merrifield} M.~R.,  2011, \mn@doi [\mnras] {10.1111/j.1365-2966.2011.18752.x}, \href {https://ui.adsabs.harvard.edu/abs/2011MNRAS.415.2063B} {415, 2063}

\bibitem[\protect\citeauthoryear{{Bensby}, {Alves-Brito}, {Oey}, {Yong}  \& {Mel{\'e}ndez}}{{Bensby} et~al.}{2011}]{Bensby2011}
{Bensby} T.,  {Alves-Brito} A.,  {Oey} M.~S.,  {Yong} D.,   {Mel{\'e}ndez} J.,  2011, \mn@doi [\apjl] {10.1088/2041-8205/735/2/L46}, \href {https://ui.adsabs.harvard.edu/abs/2011ApJ...735L..46B} {735, L46}

\bibitem[\protect\citeauthoryear{{Bhattacharya} et~al.,}{{Bhattacharya} et~al.}{2019}]{Bhattacharya2019}
{Bhattacharya} S.,  et~al., 2019, \mn@doi [\aap] {10.1051/0004-6361/201935898}, \href {https://ui.adsabs.harvard.edu/abs/2019A&A...631A..56B} {631, A56}

\bibitem[\protect\citeauthoryear{{Bird}, {Kazantzidis}, {Weinberg}, {Guedes}, {Callegari}, {Mayer}  \& {Madau}}{{Bird} et~al.}{2013}]{Bird2013}
{Bird} J.~C.,  {Kazantzidis} S.,  {Weinberg} D.~H.,  {Guedes} J.,  {Callegari} S.,  {Mayer} L.,   {Madau} P.,  2013, \mn@doi [\apj] {10.1088/0004-637X/773/1/43}, \href {https://ui.adsabs.harvard.edu/abs/2013ApJ...773...43B} {773, 43}

\bibitem[\protect\citeauthoryear{{Bird}, {Loebman}, {Weinberg}, {Brooks}, {Quinn}  \& {Christensen}}{{Bird} et~al.}{2021}]{Bird2021}
{Bird} J.~C.,  {Loebman} S.~R.,  {Weinberg} D.~H.,  {Brooks} A.~M.,  {Quinn} T.~R.,   {Christensen} C.~R.,  2021, \mn@doi [\mnras] {10.1093/mnras/stab289}, \href {https://ui.adsabs.harvard.edu/abs/2021MNRAS.503.1815B} {503, 1815}

\bibitem[\protect\citeauthoryear{{Blakeslee} et~al.,}{{Blakeslee} et~al.}{2009}]{Blakeslee2009}
{Blakeslee} J.~P.,  et~al., 2009, \mn@doi [\apj] {10.1088/0004-637X/694/1/556}, \href {https://ui.adsabs.harvard.edu/abs/2009ApJ...694..556B} {694, 556}

\bibitem[\protect\citeauthoryear{{Boecker}, {Leaman}, {van de Ven}, {Norris}, {Mackereth}  \& {Crain}}{{Boecker} et~al.}{2020}]{Boecker2020}
{Boecker} A.,  {Leaman} R.,  {van de Ven} G.,  {Norris} M.~A.,  {Mackereth} J.~T.,   {Crain} R.~A.,  2020, \mn@doi [\mnras] {10.1093/mnras/stz3077}, \href {https://ui.adsabs.harvard.edu/abs/2020MNRAS.491..823B} {491, 823}

\bibitem[\protect\citeauthoryear{{Bournaud}, {Elmegreen}  \& {Martig}}{{Bournaud} et~al.}{2009}]{Bournaud2009}
{Bournaud} F.,  {Elmegreen} B.~G.,   {Martig} M.,  2009, \mn@doi [\apjl] {10.1088/0004-637X/707/1/L1}, \href {https://ui.adsabs.harvard.edu/abs/2009ApJ...707L...1B} {707, L1}

\bibitem[\protect\citeauthoryear{{Bovy}, {Rix}, {Liu}, {Hogg}, {Beers}  \& {Lee}}{{Bovy} et~al.}{2012}]{Bovy2012b}
{Bovy} J.,  {Rix} H.-W.,  {Liu} C.,  {Hogg} D.~W.,  {Beers} T.~C.,   {Lee} Y.~S.,  2012, \mn@doi [\apj] {10.1088/0004-637X/753/2/148}, \href {https://ui.adsabs.harvard.edu/abs/2012ApJ...753..148B} {753, 148}

\bibitem[\protect\citeauthoryear{{Bovy}, {Rix}, {Schlafly}, {Nidever}, {Holtzman}, {Shetrone}  \& {Beers}}{{Bovy} et~al.}{2016}]{Bovy2016}
{Bovy} J.,  {Rix} H.-W.,  {Schlafly} E.~F.,  {Nidever} D.~L.,  {Holtzman} J.~A.,  {Shetrone} M.,   {Beers} T.~C.,  2016, \mn@doi [\apj] {10.3847/0004-637X/823/1/30}, \href {https://ui.adsabs.harvard.edu/abs/2016ApJ...823...30B} {823, 30}

\bibitem[\protect\citeauthoryear{{Brook}, {Kawata}, {Gibson}  \& {Freeman}}{{Brook} et~al.}{2004}]{Brook2004}
{Brook} C.~B.,  {Kawata} D.,  {Gibson} B.~K.,   {Freeman} K.~C.,  2004, \mn@doi [\apj] {10.1086/422709}, \href {https://ui.adsabs.harvard.edu/abs/2004ApJ...612..894B} {612, 894}

\bibitem[\protect\citeauthoryear{{Brook}, {Stinson}, {Gibson}, {Ro{\v{s}}kar}, {Wadsley}  \& {Quinn}}{{Brook} et~al.}{2012}]{Brook2012}
{Brook} C.~B.,  {Stinson} G.,  {Gibson} B.~K.,  {Ro{\v{s}}kar} R.,  {Wadsley} J.,   {Quinn} T.,  2012, \mn@doi [\mnras] {10.1111/j.1365-2966.2011.19740.x}, \href {https://ui.adsabs.harvard.edu/abs/2012MNRAS.419..771B} {419, 771}

\bibitem[\protect\citeauthoryear{{Buck}, {Obreja}, {Macci{\`o}}, {Minchev}, {Dutton}  \& {Ostriker}}{{Buck} et~al.}{2020}]{Buck2020}
{Buck} T.,  {Obreja} A.,  {Macci{\`o}} A.~V.,  {Minchev} I.,  {Dutton} A.~A.,   {Ostriker} J.~P.,  2020, \mn@doi [\mnras] {10.1093/mnras/stz3241}, \href {https://ui.adsabs.harvard.edu/abs/2020MNRAS.491.3461B} {491, 3461}

\bibitem[\protect\citeauthoryear{{Bureau}, {Aronica}, {Athanassoula}, {Dettmar}, {Bosma}  \& {Freeman}}{{Bureau} et~al.}{2006}]{Bureau2006}
{Bureau} M.,  {Aronica} G.,  {Athanassoula} E.,  {Dettmar} R.~J.,  {Bosma} A.,   {Freeman} K.~C.,  2006, \mn@doi [\mnras] {10.1111/j.1365-2966.2006.10471.x}, \href {https://ui.adsabs.harvard.edu/abs/2006MNRAS.370..753B} {370, 753}

\bibitem[\protect\citeauthoryear{{Cadiou}, {Pontzen}  \& {Peiris}}{{Cadiou} et~al.}{2022}]{Cadiou2022}
{Cadiou} C.,  {Pontzen} A.,   {Peiris} H.~V.,  2022, \mn@doi [\mnras] {10.1093/mnras/stac2858}, \href {https://ui.adsabs.harvard.edu/abs/2022MNRAS.517.3459C} {517, 3459}

\bibitem[\protect\citeauthoryear{{Cappellari}}{{Cappellari}}{2017}]{Cappellari2017}
{Cappellari} M.,  2017, \mn@doi [\mnras] {10.1093/mnras/stw3020}, \href {https://ui.adsabs.harvard.edu/abs/2017MNRAS.466..798C} {466, 798}

\bibitem[\protect\citeauthoryear{{Cappellari} \& {Copin}}{{Cappellari} \& {Copin}}{2003}]{Cappellari2003}
{Cappellari} M.,  {Copin} Y.,  2003, \mn@doi [\mnras] {10.1046/j.1365-8711.2003.06541.x}, \href {https://ui.adsabs.harvard.edu/abs/2003MNRAS.342..345C} {342, 345}

\bibitem[\protect\citeauthoryear{{Cappellari} \& {Emsellem}}{{Cappellari} \& {Emsellem}}{2004}]{Cappellari2004}
{Cappellari} M.,  {Emsellem} E.,  2004, \mn@doi [\pasp] {10.1086/381875}, \href {https://ui.adsabs.harvard.edu/abs/2004PASP..116..138C} {116, 138}

\bibitem[\protect\citeauthoryear{{Carlberg} \& {Sellwood}}{{Carlberg} \& {Sellwood}}{1985}]{Carlberg1985}
{Carlberg} R.~G.,  {Sellwood} J.~A.,  1985, \mn@doi [\apj] {10.1086/163134}, \href {https://ui.adsabs.harvard.edu/abs/1985ApJ...292...79C} {292, 79}

\bibitem[\protect\citeauthoryear{{Casagrande} et~al.,}{{Casagrande} et~al.}{2016}]{Casagrande2016}
{Casagrande} L.,  et~al., 2016, \mn@doi [\mnras] {10.1093/mnras/stv2320}, \href {https://ui.adsabs.harvard.edu/abs/2016MNRAS.455..987C} {455, 987}

\bibitem[\protect\citeauthoryear{{Cebri{\'a}n} \& {Trujillo}}{{Cebri{\'a}n} \& {Trujillo}}{2014}]{Cebrian2014}
{Cebri{\'a}n} M.,  {Trujillo} I.,  2014, \mn@doi [\mnras] {10.1093/mnras/stu1375}, \href {https://ui.adsabs.harvard.edu/abs/2014MNRAS.444..682C} {444, 682}

\bibitem[\protect\citeauthoryear{{Ceverino}, {Primack}, {Dekel}  \& {Kassin}}{{Ceverino} et~al.}{2017}]{Ceverino2017}
{Ceverino} D.,  {Primack} J.,  {Dekel} A.,   {Kassin} S.~A.,  2017, \mn@doi [\mnras] {10.1093/mnras/stx289}, \href {https://ui.adsabs.harvard.edu/abs/2017MNRAS.467.2664C} {467, 2664}

\bibitem[\protect\citeauthoryear{{Cheng} et~al.,}{{Cheng} et~al.}{2012}]{Cheng2012b}
{Cheng} J.~Y.,  et~al., 2012, \mn@doi [\apj] {10.1088/0004-637X/752/1/51}, \href {https://ui.adsabs.harvard.edu/abs/2012ApJ...752...51C} {752, 51}

\bibitem[\protect\citeauthoryear{{Chiang}, {Overzier}  \& {Gebhardt}}{{Chiang} et~al.}{2013}]{Chiang2013}
{Chiang} Y.-K.,  {Overzier} R.,   {Gebhardt} K.,  2013, \mn@doi [\apj] {10.1088/0004-637X/779/2/127}, \href {https://ui.adsabs.harvard.edu/abs/2013ApJ...779..127C} {779, 127}

\bibitem[\protect\citeauthoryear{{Cid Fernandes} et~al.,}{{Cid Fernandes} et~al.}{2013}]{CidFernandes2013}
{Cid Fernandes} R.,  et~al., 2013, \mn@doi [\aap] {10.1051/0004-6361/201220616}, \href {https://ui.adsabs.harvard.edu/abs/2013A&A...557A..86C} {557, A86}

\bibitem[\protect\citeauthoryear{{Comer{\'o}n}, {Salo}, {Janz}, {Laurikainen}  \& {Yoachim}}{{Comer{\'o}n} et~al.}{2015}]{Comeron2015}
{Comer{\'o}n} S.,  {Salo} H.,  {Janz} J.,  {Laurikainen} E.,   {Yoachim} P.,  2015, \mn@doi [\aap] {10.1051/0004-6361/201526815}, \href {https://ui.adsabs.harvard.edu/abs/2015A&A...584A..34C} {584, A34}

\bibitem[\protect\citeauthoryear{{Comer{\'o}n}, {Salo}, {Peletier}  \& {Mentz}}{{Comer{\'o}n} et~al.}{2016}]{Comeron2016}
{Comer{\'o}n} S.,  {Salo} H.,  {Peletier} R.~F.,   {Mentz} J.,  2016, \mn@doi [\aap] {10.1051/0004-6361/201629292}, \href {https://ui.adsabs.harvard.edu/abs/2016A&A...593L...6C} {593, L6}

\bibitem[\protect\citeauthoryear{{Comer{\'o}n}, {Salo}  \& {Knapen}}{{Comer{\'o}n} et~al.}{2018}]{Comeron2018}
{Comer{\'o}n} S.,  {Salo} H.,   {Knapen} J.~H.,  2018, \mn@doi [\aap] {10.1051/0004-6361/201731415}, \href {https://ui.adsabs.harvard.edu/abs/2018A&A...610A...5C} {610, A5}

\bibitem[\protect\citeauthoryear{{Danovich}, {Dekel}, {Hahn}, {Ceverino}  \& {Primack}}{{Danovich} et~al.}{2015}]{Danovich2015}
{Danovich} M.,  {Dekel} A.,  {Hahn} O.,  {Ceverino} D.,   {Primack} J.,  2015, \mn@doi [\mnras] {10.1093/mnras/stv270}, \href {https://ui.adsabs.harvard.edu/abs/2015MNRAS.449.2087D} {449, 2087}

\bibitem[\protect\citeauthoryear{{Davison} et~al.,}{{Davison} et~al.}{2021}]{Davison2021}
{Davison} T.~A.,  et~al., 2021, \mn@doi [\mnras] {10.1093/mnras/stab162}, \href {https://ui.adsabs.harvard.edu/abs/2021MNRAS.502.2296D} {502, 2296}

\bibitem[\protect\citeauthoryear{{Ding} et~al.,}{{Ding} et~al.}{2023}]{Ding2023}
{Ding} Y.,  et~al., 2023, \mn@doi [\aap] {10.1051/0004-6361/202244558}, \href {https://ui.adsabs.harvard.edu/abs/2023A&A...672A..84D} {672, A84}

\bibitem[\protect\citeauthoryear{{Ding}, {Zhu}, {Pillepich}, {van de Ven}, {Corsini}, {Iodice}  \& {Pinna}}{{Ding} et~al.}{2024}]{Ding2024}
{Ding} Y.,  {Zhu} L.,  {Pillepich} A.,  {van de Ven} G.,  {Corsini} E.~M.,  {Iodice} E.,   {Pinna} F.,  2024, \mn@doi [\aap] {10.1051/0004-6361/202449380}, \href {https://ui.adsabs.harvard.edu/abs/2024A&A...686A.184D} {686, A184}

\bibitem[\protect\citeauthoryear{{Dorman} et~al.,}{{Dorman} et~al.}{2015}]{Dorman2015}
{Dorman} C.~E.,  et~al., 2015, \mn@doi [\apj] {10.1088/0004-637X/803/1/24}, \href {https://ui.adsabs.harvard.edu/abs/2015ApJ...803...24D} {803, 24}

\bibitem[\protect\citeauthoryear{{Drinkwater}, {Gregg}  \& {Colless}}{{Drinkwater} et~al.}{2001}]{Drinkwater2001}
{Drinkwater} M.~J.,  {Gregg} M.~D.,   {Colless} M.,  2001, \mn@doi [\apjl] {10.1086/319113}, \href {https://ui.adsabs.harvard.edu/abs/2001ApJ...548L.139D} {548, L139}

\bibitem[\protect\citeauthoryear{{Dubois} et~al.,}{{Dubois} et~al.}{2021}]{Dubois2021}
{Dubois} Y.,  et~al., 2021, \mn@doi [\aap] {10.1051/0004-6361/202039429}, \href {https://ui.adsabs.harvard.edu/abs/2021A&A...651A.109D} {651, A109}

\bibitem[\protect\citeauthoryear{{Dutton} \& {van den Bosch}}{{Dutton} \& {van den Bosch}}{2009}]{Dutton2009}
{Dutton} A.~A.,  {van den Bosch} F.~C.,  2009, \mn@doi [\mnras] {10.1111/j.1365-2966.2009.14742.x}, \href {https://ui.adsabs.harvard.edu/abs/2009MNRAS.396..141D} {396, 141}

\bibitem[\protect\citeauthoryear{{Eigenbrot} \& {Bershady}}{{Eigenbrot} \& {Bershady}}{2018}]{Eigenbrot2018}
{Eigenbrot} A.,  {Bershady} M.~A.,  2018, \mn@doi [\apj] {10.3847/1538-4357/aaa45d}, \href {https://ui.adsabs.harvard.edu/abs/2018ApJ...853..114E} {853, 114}

\bibitem[\protect\citeauthoryear{{Frankel}, {Sanders}, {Rix}, {Ting}  \& {Ness}}{{Frankel} et~al.}{2019}]{Frankel2019}
{Frankel} N.,  {Sanders} J.,  {Rix} H.-W.,  {Ting} Y.-S.,   {Ness} M.,  2019, \mn@doi [\apj] {10.3847/1538-4357/ab4254}, \href {https://ui.adsabs.harvard.edu/abs/2019ApJ...884...99F} {884, 99}

\bibitem[\protect\citeauthoryear{{Gal{\'a}n-de Anta} et~al.,}{{Gal{\'a}n-de Anta} et~al.}{2022}]{Galan2022}
{Gal{\'a}n-de Anta} P.~M.,  et~al., 2022, \mn@doi [\mnras] {10.1093/mnras/stac3061}, \href {https://ui.adsabs.harvard.edu/abs/2022MNRAS.517.5992G} {517, 5992}

\bibitem[\protect\citeauthoryear{{Gal{\'a}n-de Anta}, {Capelo}, {Vasiliev}, {Dotti}, {Sarzi}, {Corsini}  \& {Morelli}}{{Gal{\'a}n-de Anta} et~al.}{2023}]{Galan2023}
{Gal{\'a}n-de Anta} P.~M.,  {Capelo} P.~R.,  {Vasiliev} E.,  {Dotti} M.,  {Sarzi} M.,  {Corsini} E.~M.,   {Morelli} L.,  2023, \mn@doi [\mnras] {10.1093/mnras/stad1593}, \href {https://ui.adsabs.harvard.edu/abs/2023MNRAS.523.3939G} {523, 3939}

\bibitem[\protect\citeauthoryear{{Garc{\'\i}a de la Cruz}, {Martig}, {Minchev}  \& {James}}{{Garc{\'\i}a de la Cruz} et~al.}{2021}]{Garcia2021}
{Garc{\'\i}a de la Cruz} J.,  {Martig} M.,  {Minchev} I.,   {James} P.,  2021, \mn@doi [\mnras] {10.1093/mnras/staa3906}, \href {https://ui.adsabs.harvard.edu/abs/2021MNRAS.501.5105G} {501, 5105}

\bibitem[\protect\citeauthoryear{{Gogarten} et~al.,}{{Gogarten} et~al.}{2010}]{Gogarten2010}
{Gogarten} S.~M.,  et~al., 2010, \mn@doi [\apj] {10.1088/0004-637X/712/2/858}, \href {https://ui.adsabs.harvard.edu/abs/2010ApJ...712..858G} {712, 858}

\bibitem[\protect\citeauthoryear{{Gonz{\'a}lez Delgado} et~al.,}{{Gonz{\'a}lez Delgado} et~al.}{2014}]{GonzalezDelgado2014}
{Gonz{\'a}lez Delgado} R.~M.,  et~al., 2014, \mn@doi [\aap] {10.1051/0004-6361/201322011}, \href {https://ui.adsabs.harvard.edu/abs/2014A&A...562A..47G} {562, A47}

\bibitem[\protect\citeauthoryear{{Grand}, {Springel}, {G{\'o}mez}, {Marinacci}, {Pakmor}, {Campbell}  \& {Jenkins}}{{Grand} et~al.}{2016}]{Grand2016}
{Grand} R. J.~J.,  {Springel} V.,  {G{\'o}mez} F.~A.,  {Marinacci} F.,  {Pakmor} R.,  {Campbell} D. J.~R.,   {Jenkins} A.,  2016, \mn@doi [\mnras] {10.1093/mnras/stw601}, \href {https://ui.adsabs.harvard.edu/abs/2016MNRAS.459..199G} {459, 199}

\bibitem[\protect\citeauthoryear{{Grand} et~al.,}{{Grand} et~al.}{2017}]{Grand2017}
{Grand} R. J.~J.,  et~al., 2017, \mn@doi [\mnras] {10.1093/mnras/stx071}, \href {https://ui.adsabs.harvard.edu/abs/2017MNRAS.467..179G} {467, 179}

\bibitem[\protect\citeauthoryear{{Grand} et~al.,}{{Grand} et~al.}{2019}]{Grand2019}
{Grand} R. J.~J.,  et~al., 2019, \mn@doi [\mnras] {10.1093/mnras/stz2928}, \href {https://ui.adsabs.harvard.edu/abs/2019MNRAS.490.4786G} {490, 4786}

\bibitem[\protect\citeauthoryear{{Gu{\'e}rou}, {Emsellem}, {Krajnovi{\'c}}, {McDermid}, {Contini}  \& {Weilbacher}}{{Gu{\'e}rou} et~al.}{2016}]{Guerou2016}
{Gu{\'e}rou} A.,  {Emsellem} E.,  {Krajnovi{\'c}} D.,  {McDermid} R.~M.,  {Contini} T.,   {Weilbacher} P.~M.,  2016, \mn@doi [\aap] {10.1051/0004-6361/201628743}, \href {https://ui.adsabs.harvard.edu/abs/2016A&A...591A.143G} {591, A143}

\bibitem[\protect\citeauthoryear{{Gunn} \& {Gott}}{{Gunn} \& {Gott}}{1972}]{Gunn1972}
{Gunn} J.~E.,  {Gott} III J.~R.,  1972, \mn@doi [\apj] {10.1086/151605}, \href {https://ui.adsabs.harvard.edu/abs/1972ApJ...176....1G} {176, 1}

\bibitem[\protect\citeauthoryear{{House} et~al.,}{{House} et~al.}{2011}]{House2011}
{House} E.~L.,  et~al., 2011, \mn@doi [\mnras] {10.1111/j.1365-2966.2011.18891.x}, \href {https://ui.adsabs.harvard.edu/abs/2011MNRAS.415.2652H} {415, 2652}

\bibitem[\protect\citeauthoryear{{Ibarra-Medel}, {Avila-Reese}, {S{\'a}nchez}, {Gonz{\'a}lez-Samaniego}  \& {Rodr{\'\i}guez-Puebla}}{{Ibarra-Medel} et~al.}{2019}]{Ibarra2019}
{Ibarra-Medel} H.~J.,  {Avila-Reese} V.,  {S{\'a}nchez} S.~F.,  {Gonz{\'a}lez-Samaniego} A.,   {Rodr{\'\i}guez-Puebla} A.,  2019, \mn@doi [\mnras] {10.1093/mnras/sty3256}, \href {https://ui.adsabs.harvard.edu/abs/2019MNRAS.483.4525I} {483, 4525}

\bibitem[\protect\citeauthoryear{{Imig} et~al.,}{{Imig} et~al.}{2023}]{Imig2023}
{Imig} J.,  et~al., 2023, \mn@doi [\apj] {10.3847/1538-4357/ace9b8}, \href {https://ui.adsabs.harvard.edu/abs/2023ApJ...954..124I} {954, 124}

\bibitem[\protect\citeauthoryear{{Imig} et~al.,}{{Imig} et~al.}{2025}]{Imig2025}
{Imig} J.,  et~al., 2025, \mn@doi [\apj] {10.3847/1538-4357/adf723}, \href {https://ui.adsabs.harvard.edu/abs/2025ApJ...990..203I} {990, 203}

\bibitem[\protect\citeauthoryear{{Iodice} et~al.,}{{Iodice} et~al.}{2019a}]{Iodice2019b}
{Iodice} E.,  et~al., 2019a, \mn@doi [\aap] {10.1051/0004-6361/201833741}, \href {https://ui.adsabs.harvard.edu/abs/2019A&A...623A...1I} {623, A1}

\bibitem[\protect\citeauthoryear{{Iodice} et~al.,}{{Iodice} et~al.}{2019b}]{Iodice2019a}
{Iodice} E.,  et~al., 2019b, \mn@doi [\aap] {10.1051/0004-6361/201935721}, \href {https://ui.adsabs.harvard.edu/abs/2019A&A...627A.136I} {627, A136}

\bibitem[\protect\citeauthoryear{{Joshi}, {Pillepich}, {Nelson}, {Marinacci}, {Springel}, {Rodriguez-Gomez}, {Vogelsberger}  \& {Hernquist}}{{Joshi} et~al.}{2020}]{Joshi2020}
{Joshi} G.~D.,  {Pillepich} A.,  {Nelson} D.,  {Marinacci} F.,  {Springel} V.,  {Rodriguez-Gomez} V.,  {Vogelsberger} M.,   {Hernquist} L.,  2020, \mn@doi [\mnras] {10.1093/mnras/staa1668}, \href {https://ui.adsabs.harvard.edu/abs/2020MNRAS.496.2673J} {496, 2673}

\bibitem[\protect\citeauthoryear{{Kasparova}, {Katkov}, {Chilingarian}, {Silchenko}, {Moiseev}  \& {Borisov}}{{Kasparova} et~al.}{2016}]{Kasparova2016}
{Kasparova} A.~V.,  {Katkov} I.~Y.,  {Chilingarian} I.~V.,  {Silchenko} O.~K.,  {Moiseev} A.~V.,   {Borisov} S.~B.,  2016, \mn@doi [\mnras] {10.1093/mnrasl/slw083}, \href {https://ui.adsabs.harvard.edu/abs/2016MNRAS.460L..89K} {460, L89}

\bibitem[\protect\citeauthoryear{{Kasparova}, {Katkov}  \& {Chilingarian}}{{Kasparova} et~al.}{2020}]{Kasparova2020}
{Kasparova} A.~V.,  {Katkov} I.~Y.,   {Chilingarian} I.~V.,  2020, \mn@doi [\mnras] {10.1093/mnras/staa611}, \href {https://ui.adsabs.harvard.edu/abs/2020MNRAS.493.5464K} {493, 5464}

\bibitem[\protect\citeauthoryear{{Kassin}, {Brooks}, {Governato}, {Weiner}  \& {Gardner}}{{Kassin} et~al.}{2014}]{Kassin2014}
{Kassin} S.~A.,  {Brooks} A.,  {Governato} F.,  {Weiner} B.~J.,   {Gardner} J.~P.,  2014, \mn@doi [\apj] {10.1088/0004-637X/790/2/89}, \href {https://ui.adsabs.harvard.edu/abs/2014ApJ...790...89K} {790, 89}

\bibitem[\protect\citeauthoryear{{Khoperskov} et~al.,}{{Khoperskov} et~al.}{2025}]{Khoperskov2025}
{Khoperskov} S.,  et~al., 2025, \mn@doi [\aap] {10.1051/0004-6361/202453305}, \href {https://ui.adsabs.harvard.edu/abs/2025A&A...700A..89K} {700, A89}

\bibitem[\protect\citeauthoryear{{Kleiner} et~al.,}{{Kleiner} et~al.}{2025}]{Kleiner2025}
{Kleiner} D.,  et~al., 2025, arXiv e-prints, \href {https://ui.adsabs.harvard.edu/abs/2025arXiv251115795K} {p. arXiv:2511.15795}

\bibitem[\protect\citeauthoryear{{Lagos} et~al.,}{{Lagos} et~al.}{2018}]{Lagos2018}
{Lagos} C. d.~P.,  et~al., 2018, \mn@doi [\mnras] {10.1093/mnras/stx2667}, \href {https://ui.adsabs.harvard.edu/abs/2018MNRAS.473.4956L} {473, 4956}

\bibitem[\protect\citeauthoryear{{Larson}}{{Larson}}{1976}]{Larson1976}
{Larson} R.~B.,  1976, \mn@doi [\mnras] {10.1093/mnras/176.1.31}, \href {https://ui.adsabs.harvard.edu/abs/1976MNRAS.176...31L} {176, 31}

\bibitem[\protect\citeauthoryear{{Larson}, {Tinsley}  \& {Caldwell}}{{Larson} et~al.}{1980}]{Larson1980}
{Larson} R.~B.,  {Tinsley} B.~M.,   {Caldwell} C.~N.,  1980, \mn@doi [\apj] {10.1086/157917}, \href {https://ui.adsabs.harvard.edu/abs/1980ApJ...237..692L} {237, 692}

\bibitem[\protect\citeauthoryear{{Leaman} et~al.,}{{Leaman} et~al.}{2017}]{Leaman2017}
{Leaman} R.,  et~al., 2017, \mn@doi [\mnras] {10.1093/mnras/stx2014}, \href {https://ui.adsabs.harvard.edu/abs/2017MNRAS.472.1879L} {472, 1879}

\bibitem[\protect\citeauthoryear{{Loni} et~al.,}{{Loni} et~al.}{2021}]{Loni2021}
{Loni} A.,  et~al., 2021, \mn@doi [\aap] {10.1051/0004-6361/202039803}, \href {https://ui.adsabs.harvard.edu/abs/2021A&A...648A..31L} {648, A31}

\bibitem[\protect\citeauthoryear{{L{\"u}tticke}, {Dettmar}  \& {Pohlen}}{{L{\"u}tticke} et~al.}{2000}]{Lutticke2000}
{L{\"u}tticke} R.,  {Dettmar} R.~J.,   {Pohlen} M.,  2000, \mn@doi [\aaps] {10.1051/aas:2000354}, \href {https://ui.adsabs.harvard.edu/abs/2000A&AS..145..405L} {145, 405}

\bibitem[\protect\citeauthoryear{{Ma}, {Hopkins}, {Wetzel}, {Kirby}, {Angl{\'e}s-Alc{\'a}zar}, {Faucher-Gigu{\`e}re}, {Kere{\v{s}}}  \& {Quataert}}{{Ma} et~al.}{2017}]{Ma2017}
{Ma} X.,  {Hopkins} P.~F.,  {Wetzel} A.~R.,  {Kirby} E.~N.,  {Angl{\'e}s-Alc{\'a}zar} D.,  {Faucher-Gigu{\`e}re} C.-A.,  {Kere{\v{s}}} D.,   {Quataert} E.,  2017, \mn@doi [\mnras] {10.1093/mnras/stx273}, \href {https://ui.adsabs.harvard.edu/abs/2017MNRAS.467.2430M} {467, 2430}

\bibitem[\protect\citeauthoryear{{Mackereth} et~al.,}{{Mackereth} et~al.}{2017}]{Mackereth2017}
{Mackereth} J.~T.,  et~al., 2017, \mn@doi [\mnras] {10.1093/mnras/stx1774}, \href {https://ui.adsabs.harvard.edu/abs/2017MNRAS.471.3057M} {471, 3057}

\bibitem[\protect\citeauthoryear{{Maller} \& {Dekel}}{{Maller} \& {Dekel}}{2002}]{Maller2002}
{Maller} A.~H.,  {Dekel} A.,  2002, \mn@doi [\mnras] {10.1046/j.1365-8711.2002.05646.x}, \href {https://ui.adsabs.harvard.edu/abs/2002MNRAS.335..487M} {335, 487}

\bibitem[\protect\citeauthoryear{{Martig}, {Minchev}  \& {Flynn}}{{Martig} et~al.}{2014a}]{Martig2014a}
{Martig} M.,  {Minchev} I.,   {Flynn} C.,  2014a, \mn@doi [\mnras] {10.1093/mnras/stu1003}, \href {https://ui.adsabs.harvard.edu/abs/2014MNRAS.442.2474M} {442, 2474}

\bibitem[\protect\citeauthoryear{{Martig}, {Minchev}  \& {Flynn}}{{Martig} et~al.}{2014b}]{Martig2014b}
{Martig} M.,  {Minchev} I.,   {Flynn} C.,  2014b, \mn@doi [\mnras] {10.1093/mnras/stu1322}, \href {https://ui.adsabs.harvard.edu/abs/2014MNRAS.443.2452M} {443, 2452}

\bibitem[\protect\citeauthoryear{{Martig}, {Minchev}, {Ness}, {Fouesneau}  \& {Rix}}{{Martig} et~al.}{2016}]{Martig2016}
{Martig} M.,  {Minchev} I.,  {Ness} M.,  {Fouesneau} M.,   {Rix} H.-W.,  2016, \mn@doi [\apj] {10.3847/0004-637X/831/2/139}, \href {https://ui.adsabs.harvard.edu/abs/2016ApJ...831..139M} {831, 139}

\bibitem[\protect\citeauthoryear{{Martig} et~al.,}{{Martig} et~al.}{2021}]{Martig2021}
{Martig} M.,  et~al., 2021, \mn@doi [\mnras] {10.1093/mnras/stab2729}, \href {https://ui.adsabs.harvard.edu/abs/2021MNRAS.508.2458M} {508, 2458}

\bibitem[\protect\citeauthoryear{{Mart{\'\i}n-Navarro}, {van de Ven}  \& {Y{\i}ld{\i}r{\i}m}}{{Mart{\'\i}n-Navarro} et~al.}{2019}]{MartinNavarro2019}
{Mart{\'\i}n-Navarro} I.,  {van de Ven} G.,   {Y{\i}ld{\i}r{\i}m} A.,  2019, \mn@doi [\mnras] {10.1093/mnras/stz1544}, \href {https://ui.adsabs.harvard.edu/abs/2019MNRAS.487.4939M} {487, 4939}

\bibitem[\protect\citeauthoryear{{Mart{\'\i}n-Navarro} et~al.,}{{Mart{\'\i}n-Navarro} et~al.}{2021}]{MartinNavarro2021}
{Mart{\'\i}n-Navarro} I.,  et~al., 2021, \mn@doi [\aap] {10.1051/0004-6361/202141348}, \href {https://ui.adsabs.harvard.edu/abs/2021A&A...654A..59M} {654, A59}

\bibitem[\protect\citeauthoryear{{McCluskey}, {Wetzel}, {Loebman}, {Moreno}, {Faucher-Gigu{\`e}re}  \& {Hopkins}}{{McCluskey} et~al.}{2024}]{McCluskey2024}
{McCluskey} F.,  {Wetzel} A.,  {Loebman} S.~R.,  {Moreno} J.,  {Faucher-Gigu{\`e}re} C.-A.,   {Hopkins} P.~F.,  2024, \mn@doi [\mnras] {10.1093/mnras/stad3547}, \href {https://ui.adsabs.harvard.edu/abs/2024MNRAS.527.6926M} {527, 6926}

\bibitem[\protect\citeauthoryear{{Meng} \& {Gnedin}}{{Meng} \& {Gnedin}}{2021}]{Meng2021}
{Meng} X.,  {Gnedin} O.~Y.,  2021, \mn@doi [\mnras] {10.1093/mnras/stab088}, \href {https://ui.adsabs.harvard.edu/abs/2021MNRAS.502.1433M} {502, 1433}

\bibitem[\protect\citeauthoryear{{Minchev}, {Famaey}, {Quillen}, {Dehnen}, {Martig}  \& {Siebert}}{{Minchev} et~al.}{2012}]{Minchev2012}
{Minchev} I.,  {Famaey} B.,  {Quillen} A.~C.,  {Dehnen} W.,  {Martig} M.,   {Siebert} A.,  2012, \mn@doi [\aap] {10.1051/0004-6361/201219714}, \href {https://ui.adsabs.harvard.edu/abs/2012A&A...548A.127M} {548, A127}

\bibitem[\protect\citeauthoryear{{Minchev}, {Martig}, {Streich}, {Scannapieco}, {de Jong}  \& {Steinmetz}}{{Minchev} et~al.}{2015}]{Minchev2015}
{Minchev} I.,  {Martig} M.,  {Streich} D.,  {Scannapieco} C.,  {de Jong} R.~S.,   {Steinmetz} M.,  2015, \mn@doi [\apjl] {10.1088/2041-8205/804/1/L9}, \href {https://ui.adsabs.harvard.edu/abs/2015ApJ...804L...9M} {804, L9}

\bibitem[\protect\citeauthoryear{{Minchev}, {Steinmetz}, {Chiappini}, {Martig}, {Anders}, {Matijevic}  \& {de Jong}}{{Minchev} et~al.}{2017}]{Minchev2017}
{Minchev} I.,  {Steinmetz} M.,  {Chiappini} C.,  {Martig} M.,  {Anders} F.,  {Matijevic} G.,   {de Jong} R.~S.,  2017, \mn@doi [\apj] {10.3847/1538-4357/834/1/27}, \href {https://ui.adsabs.harvard.edu/abs/2017ApJ...834...27M} {834, 27}

\bibitem[\protect\citeauthoryear{{Mo}, {Mao}  \& {White}}{{Mo} et~al.}{1998}]{Mo1998}
{Mo} H.~J.,  {Mao} S.,   {White} S. D.~M.,  1998, \mn@doi [\mnras] {10.1046/j.1365-8711.1998.01227.x}, \href {https://ui.adsabs.harvard.edu/abs/1998MNRAS.295..319M} {295, 319}

\bibitem[\protect\citeauthoryear{{Moore}, {Lake}  \& {Katz}}{{Moore} et~al.}{1998}]{Moore1998}
{Moore} B.,  {Lake} G.,   {Katz} N.,  1998, \mn@doi [\apj] {10.1086/305264}, \href {https://ui.adsabs.harvard.edu/abs/1998ApJ...495..139M} {495, 139}

\bibitem[\protect\citeauthoryear{{Mould}}{{Mould}}{2005}]{Mould2005}
{Mould} J.,  2005, \mn@doi [\aj] {10.1086/427248}, \href {https://ui.adsabs.harvard.edu/abs/2005AJ....129..698M} {129, 698}

\bibitem[\protect\citeauthoryear{{Nordstr{\"o}m} et~al.,}{{Nordstr{\"o}m} et~al.}{2004}]{Nordstrom2004}
{Nordstr{\"o}m} B.,  et~al., 2004, \mn@doi [\aap] {10.1051/0004-6361:20035959}, \href {https://ui.adsabs.harvard.edu/abs/2004A&A...418..989N} {418, 989}

\bibitem[\protect\citeauthoryear{{Ocvirk}, {Pichon}, {Lan{\c{c}}on}  \& {Thi{\'e}baut}}{{Ocvirk} et~al.}{2006}]{Ocvirk2006}
{Ocvirk} P.,  {Pichon} C.,  {Lan{\c{c}}on} A.,   {Thi{\'e}baut} E.,  2006, \mn@doi [\mnras] {10.1111/j.1365-2966.2005.09323.x}, \href {https://ui.adsabs.harvard.edu/abs/2006MNRAS.365...74O} {365, 74}

\bibitem[\protect\citeauthoryear{{Parikh}, {Thomas}, {Maraston}, {Westfall}, {Andrews}, {Boardman}, {Drory}  \& {Oyarzun}}{{Parikh} et~al.}{2021}]{Parikh2021}
{Parikh} T.,  {Thomas} D.,  {Maraston} C.,  {Westfall} K.~B.,  {Andrews} B.~H.,  {Boardman} N.~F.,  {Drory} N.,   {Oyarzun} G.,  2021, \mn@doi [\mnras] {10.1093/mnras/stab449}, \href {https://ui.adsabs.harvard.edu/abs/2021MNRAS.502.5508P} {502, 5508}

\bibitem[\protect\citeauthoryear{{Park} et~al.,}{{Park} et~al.}{2021}]{Park2021}
{Park} M.~J.,  et~al., 2021, \mn@doi [\apjs] {10.3847/1538-4365/abe937}, \href {https://ui.adsabs.harvard.edu/abs/2021ApJS..254....2P} {254, 2}

\bibitem[\protect\citeauthoryear{{Pessa} et~al.,}{{Pessa} et~al.}{2023}]{Pessa2023}
{Pessa} I.,  et~al., 2023, \mn@doi [\aap] {10.1051/0004-6361/202245673}, \href {https://ui.adsabs.harvard.edu/abs/2023A&A...673A.147P} {673, A147}

\bibitem[\protect\citeauthoryear{{Peterken}, {Merrifield}, {Arag{\'o}n-Salamanca}, {Fraser-McKelvie}, {Avila-Reese}, {Riffel}, {Knapen}  \& {Drory}}{{Peterken} et~al.}{2020}]{Peterken2020}
{Peterken} T.,  {Merrifield} M.,  {Arag{\'o}n-Salamanca} A.,  {Fraser-McKelvie} A.,  {Avila-Reese} V.,  {Riffel} R.,  {Knapen} J.,   {Drory} N.,  2020, \mn@doi [\mnras] {10.1093/mnras/staa1303}, \href {https://ui.adsabs.harvard.edu/abs/2020MNRAS.495.3387P} {495, 3387}

\bibitem[\protect\citeauthoryear{{Pichon}, {Pogosyan}, {Kimm}, {Slyz}, {Devriendt}  \& {Dubois}}{{Pichon} et~al.}{2011}]{Pichon2011}
{Pichon} C.,  {Pogosyan} D.,  {Kimm} T.,  {Slyz} A.,  {Devriendt} J.,   {Dubois} Y.,  2011, \mn@doi [\mnras] {10.1111/j.1365-2966.2011.19640.x}, \href {https://ui.adsabs.harvard.edu/abs/2011MNRAS.418.2493P} {418, 2493}

\bibitem[\protect\citeauthoryear{{Pinna}, {Falc{\'o}n-Barroso}, {Martig}, {Mart{\'\i}nez-Valpuesta}, {M{\'e}ndez-Abreu}, {van de Ven}, {Leaman}  \& {Lyubenova}}{{Pinna} et~al.}{2018}]{Pinna2018}
{Pinna} F.,  {Falc{\'o}n-Barroso} J.,  {Martig} M.,  {Mart{\'\i}nez-Valpuesta} I.,  {M{\'e}ndez-Abreu} J.,  {van de Ven} G.,  {Leaman} R.,   {Lyubenova} M.,  2018, \mn@doi [\mnras] {10.1093/mnras/stx3331}, \href {https://ui.adsabs.harvard.edu/abs/2018MNRAS.475.2697P} {475, 2697}

\bibitem[\protect\citeauthoryear{{Pinna} et~al.,}{{Pinna} et~al.}{2019a}]{Pinna2019a}
{Pinna} F.,  et~al., 2019a, \mn@doi [\aap] {10.1051/0004-6361/201833193}, \href {https://ui.adsabs.harvard.edu/abs/2019A&A...623A..19P} {623, A19}

\bibitem[\protect\citeauthoryear{{Pinna} et~al.,}{{Pinna} et~al.}{2019b}]{Pinna2019b}
{Pinna} F.,  et~al., 2019b, \mn@doi [\aap] {10.1051/0004-6361/201935154}, \href {https://ui.adsabs.harvard.edu/abs/2019A&A...625A..95P} {625, A95}

\bibitem[\protect\citeauthoryear{{Poci}, {McDermid}, {Zhu}  \& {van de Ven}}{{Poci} et~al.}{2019}]{Poci2019}
{Poci} A.,  {McDermid} R.~M.,  {Zhu} L.,   {van de Ven} G.,  2019, \mn@doi [\mnras] {10.1093/mnras/stz1154}, \href {https://ui.adsabs.harvard.edu/abs/2019MNRAS.487.3776P} {487, 3776}

\bibitem[\protect\citeauthoryear{{Poci} et~al.,}{{Poci} et~al.}{2021}]{Poci2021}
{Poci} A.,  et~al., 2021, \mn@doi [\aap] {10.1051/0004-6361/202039644}, \href {https://ui.adsabs.harvard.edu/abs/2021A&A...647A.145P} {647, A145}

\bibitem[\protect\citeauthoryear{{Qu}, {Di Matteo}, {Lehnert}  \& {van Driel}}{{Qu} et~al.}{2011}]{Qu2011}
{Qu} Y.,  {Di Matteo} P.,  {Lehnert} M.~D.,   {van Driel} W.,  2011, \mn@doi [\aap] {10.1051/0004-6361/201015224}, \href {https://ui.adsabs.harvard.edu/abs/2011A&A...530A..10Q} {530, A10}

\bibitem[\protect\citeauthoryear{{Quillen} \& {Garnett}}{{Quillen} \& {Garnett}}{2001}]{Quillen2001}
{Quillen} A.~C.,  {Garnett} D.~R.,  2001, in {Funes} J.~G.,  {Corsini} E.~M.,  eds,  Astronomical Society of the Pacific Conference Series Vol. 230, Galaxy Disks and Disk Galaxies. pp 87--88

\bibitem[\protect\citeauthoryear{{Quinn}, {Hernquist}  \& {Fullagar}}{{Quinn} et~al.}{1993}]{Quinn1993}
{Quinn} P.~J.,  {Hernquist} L.,   {Fullagar} D.~P.,  1993, \mn@doi [\apj] {10.1086/172184}, \href {https://ui.adsabs.harvard.edu/abs/1993ApJ...403...74Q} {403, 74}

\bibitem[\protect\citeauthoryear{{Rejkuba}, {Mouhcine}  \& {Ibata}}{{Rejkuba} et~al.}{2009}]{Rejkuba2009}
{Rejkuba} M.,  {Mouhcine} M.,   {Ibata} R.,  2009, \mn@doi [\mnras] {10.1111/j.1365-2966.2009.14821.x}, \href {https://ui.adsabs.harvard.edu/abs/2009MNRAS.396.1231R} {396, 1231}

\bibitem[\protect\citeauthoryear{{Sacchi} et~al.,}{{Sacchi} et~al.}{2019}]{Sacchi2019}
{Sacchi} E.,  et~al., 2019, \mn@doi [\apj] {10.3847/1538-4357/ab1de1}, \href {https://ui.adsabs.harvard.edu/abs/2019ApJ...878....1S} {878, 1}

\bibitem[\protect\citeauthoryear{{S{\'a}nchez-Bl{\'a}zquez} et~al.,}{{S{\'a}nchez-Bl{\'a}zquez} et~al.}{2014}]{SanchezBlazquez2014}
{S{\'a}nchez-Bl{\'a}zquez} P.,  et~al., 2014, \mn@doi [\aap] {10.1051/0004-6361/201423635}, \href {https://ui.adsabs.harvard.edu/abs/2014A&A...570A...6S} {570, A6}

\bibitem[\protect\citeauthoryear{{Sanders} \& {Das}}{{Sanders} \& {Das}}{2018}]{Sanders2018}
{Sanders} J.~L.,  {Das} P.,  2018, \mn@doi [\mnras] {10.1093/mnras/sty2490}, \href {https://ui.adsabs.harvard.edu/abs/2018MNRAS.481.4093S} {481, 4093}

\bibitem[\protect\citeauthoryear{{Sarzi} et~al.,}{{Sarzi} et~al.}{2018}]{Sarzi2018}
{Sarzi} M.,  et~al., 2018, \mn@doi [\aap] {10.1051/0004-6361/201833137}, \href {https://ui.adsabs.harvard.edu/abs/2018A&A...616A.121S} {616, A121}

\bibitem[\protect\citeauthoryear{{Sattler}, {Pinna}, {Neumayer}, {Falc{\'o}n-Barroso}, {Martig}, {Gadotti}, {van de Ven}  \& {Minchev}}{{Sattler} et~al.}{2023}]{Sattler2023}
{Sattler} N.,  {Pinna} F.,  {Neumayer} N.,  {Falc{\'o}n-Barroso} J.,  {Martig} M.,  {Gadotti} D.~A.,  {van de Ven} G.,   {Minchev} I.,  2023, \mn@doi [\mnras] {10.1093/mnras/stad275}, \href {https://ui.adsabs.harvard.edu/abs/2023MNRAS.520.3066S} {520, 3066}

\bibitem[\protect\citeauthoryear{{Sattler}, {Pinna}, {Comer{\'o}n}, {Martig}, {Falc{\'o}n-Barroso}, {Mart{\'\i}n-Navarro}  \& {Neumayer}}{{Sattler} et~al.}{2025}]{Sattler2025}
{Sattler} N.,  {Pinna} F.,  {Comer{\'o}n} S.,  {Martig} M.,  {Falc{\'o}n-Barroso} J.,  {Mart{\'\i}n-Navarro} I.,   {Neumayer} N.,  2025, \mn@doi [\aap] {10.1051/0004-6361/202452528}, \href {https://ui.adsabs.harvard.edu/abs/2025A&A...698A.235S} {698, A235}

\bibitem[\protect\citeauthoryear{{Scott}, {van de Sande}, {Sharma}, {Bland-Hawthorn}, {Freeman}, {Gerhard}, {Hayden}  \& {McDermid}}{{Scott} et~al.}{2021}]{Scott2021}
{Scott} N.,  {van de Sande} J.,  {Sharma} S.,  {Bland-Hawthorn} J.,  {Freeman} K.,  {Gerhard} O.,  {Hayden} M.~R.,   {McDermid} R.,  2021, \mn@doi [\apjl] {10.3847/2041-8213/abfc57}, \href {https://ui.adsabs.harvard.edu/abs/2021ApJ...913L..11S} {913, L11}

\bibitem[\protect\citeauthoryear{{Sellwood} \& {Binney}}{{Sellwood} \& {Binney}}{2002}]{Sellwood2002}
{Sellwood} J.~A.,  {Binney} J.~J.,  2002, \mn@doi [\mnras] {10.1046/j.1365-8711.2002.05806.x}, \href {https://ui.adsabs.harvard.edu/abs/2002MNRAS.336..785S} {336, 785}

\bibitem[\protect\citeauthoryear{{Seth}, {Dalcanton}  \& {de Jong}}{{Seth} et~al.}{2005}]{Seth2005}
{Seth} A.~C.,  {Dalcanton} J.~J.,   {de Jong} R.~S.,  2005, \mn@doi [\aj] {10.1086/444620}, \href {https://ui.adsabs.harvard.edu/abs/2005AJ....130.1574S} {130, 1574}

\bibitem[\protect\citeauthoryear{{Sotillo-Ramos}, {Donnari}, {Pillepich}, {Frankel}, {Nelson}, {Springel}  \& {Hernquist}}{{Sotillo-Ramos} et~al.}{2023}]{Sotillo2023}
{Sotillo-Ramos} D.,  {Donnari} M.,  {Pillepich} A.,  {Frankel} N.,  {Nelson} D.,  {Springel} V.,   {Hernquist} L.,  2023, \mn@doi [\mnras] {10.1093/mnras/stad1485}, \href {https://ui.adsabs.harvard.edu/abs/2023MNRAS.523.3915S} {523, 3915}

\bibitem[\protect\citeauthoryear{{Soubiran}, {Bienaym{\'e}}, {Mishenina}  \& {Kovtyukh}}{{Soubiran} et~al.}{2008}]{Soubiran2008}
{Soubiran} C.,  {Bienaym{\'e}} O.,  {Mishenina} T.~V.,   {Kovtyukh} V.~V.,  2008, \mn@doi [\aap] {10.1051/0004-6361:20078788}, \href {https://ui.adsabs.harvard.edu/abs/2008A&A...480...91S} {480, 91}

\bibitem[\protect\citeauthoryear{{Spavone} et~al.,}{{Spavone} et~al.}{2022}]{Spavone2022}
{Spavone} M.,  et~al., 2022, \mn@doi [\aap] {10.1051/0004-6361/202243290}, \href {https://ui.adsabs.harvard.edu/abs/2022A&A...663A.135S} {663, A135}

\bibitem[\protect\citeauthoryear{{Spitzer} \& {Schwarzschild}}{{Spitzer} \& {Schwarzschild}}{1951}]{Spitzer1951}
{Spitzer} Jr. L.,  {Schwarzschild} M.,  1951, \mn@doi [\apj] {10.1086/145478}, \href {https://ui.adsabs.harvard.edu/abs/1951ApJ...114..385S} {114, 385}

\bibitem[\protect\citeauthoryear{{Spitzer} \& {Schwarzschild}}{{Spitzer} \& {Schwarzschild}}{1953}]{Spitzer1953}
{Spitzer} Jr. L.,  {Schwarzschild} M.,  1953, \mn@doi [\apj] {10.1086/145730}, \href {https://ui.adsabs.harvard.edu/abs/1953ApJ...118..106S} {118, 106}

\bibitem[\protect\citeauthoryear{{Stern} et~al.,}{{Stern} et~al.}{2021}]{Stern2021}
{Stern} J.,  et~al., 2021, \mn@doi [\apj] {10.3847/1538-4357/abd776}, \href {https://ui.adsabs.harvard.edu/abs/2021ApJ...911...88S} {911, 88}

\bibitem[\protect\citeauthoryear{{Stewart}, {Brooks}, {Bullock}, {Maller}, {Diemand}, {Wadsley}  \& {Moustakas}}{{Stewart} et~al.}{2013}]{Stewart2013}
{Stewart} K.~R.,  {Brooks} A.~M.,  {Bullock} J.~S.,  {Maller} A.~H.,  {Diemand} J.,  {Wadsley} J.,   {Moustakas} L.~A.,  2013, \mn@doi [\apj] {10.1088/0004-637X/769/1/74}, \href {https://ui.adsabs.harvard.edu/abs/2013ApJ...769...74S} {769, 74}

\bibitem[\protect\citeauthoryear{{Stinson} et~al.,}{{Stinson} et~al.}{2013}]{Stinson2013}
{Stinson} G.~S.,  et~al., 2013, \mn@doi [\mnras] {10.1093/mnras/stt1600}, \href {https://ui.adsabs.harvard.edu/abs/2013MNRAS.436..625S} {436, 625}

\bibitem[\protect\citeauthoryear{{Streich}, {de Jong}, {Bailin}, {Bell}, {Holwerda}, {Minchev}, {Monachesi}  \& {Radburn-Smith}}{{Streich} et~al.}{2016}]{Streich2016}
{Streich} D.,  {de Jong} R.~S.,  {Bailin} J.,  {Bell} E.~F.,  {Holwerda} B.~W.,  {Minchev} I.,  {Monachesi} A.,   {Radburn-Smith} D.~J.,  2016, \mn@doi [\aap] {10.1051/0004-6361/201526013}, \href {https://ui.adsabs.harvard.edu/abs/2016A&A...585A..97S} {585, A97}

\bibitem[\protect\citeauthoryear{{Su} et~al.,}{{Su} et~al.}{2017}]{Su2017}
{Su} Y.,  et~al., 2017, \mn@doi [\apj] {10.3847/1538-4357/aa989e}, \href {https://ui.adsabs.harvard.edu/abs/2017ApJ...851...69S} {851, 69}

\bibitem[\protect\citeauthoryear{{Tamfal}, {Mayer}, {Quinn}, {Babul}, {Madau}, {Capelo}, {Shen}  \& {Staub}}{{Tamfal} et~al.}{2022}]{Tamfal2022}
{Tamfal} T.,  {Mayer} L.,  {Quinn} T.~R.,  {Babul} A.,  {Madau} P.,  {Capelo} P.~R.,  {Shen} S.,   {Staub} M.,  2022, \mn@doi [\apj] {10.3847/1538-4357/ac558e}, \href {https://ui.adsabs.harvard.edu/abs/2022ApJ...928..106T} {928, 106}

\bibitem[\protect\citeauthoryear{{Tikhonov} \& {Galazutdinova}}{{Tikhonov} \& {Galazutdinova}}{2005}]{Tikhonov2005}
{Tikhonov} N.~A.,  {Galazutdinova} O.~A.,  2005, \mn@doi [Astrophysics] {10.1007/s10511-005-0021-8}, \href {https://ui.adsabs.harvard.edu/abs/2005Ap.....48..221T} {48, 221}

\bibitem[\protect\citeauthoryear{{Turner}, {C{\^o}t{\'e}}, {Ferrarese}, {Jord{\'a}n}, {Blakeslee}, {Mei}, {Peng}  \& {West}}{{Turner} et~al.}{2012}]{Turner2012}
{Turner} M.~L.,  {C{\^o}t{\'e}} P.,  {Ferrarese} L.,  {Jord{\'a}n} A.,  {Blakeslee} J.~P.,  {Mei} S.,  {Peng} E.~W.,   {West} M.~J.,  2012, \mn@doi [\apjs] {10.1088/0067-0049/203/1/5}, \href {https://ui.adsabs.harvard.edu/abs/2012ApJS..203....5T} {203, 5}

\bibitem[\protect\citeauthoryear{{Vazdekis} et~al.,}{{Vazdekis} et~al.}{2015}]{Vazdekis2015}
{Vazdekis} A.,  et~al., 2015, \mn@doi [\mnras] {10.1093/mnras/stv151}, \href {https://ui.adsabs.harvard.edu/abs/2015MNRAS.449.1177V} {449, 1177}

\bibitem[\protect\citeauthoryear{{Villalobos} \& {Helmi}}{{Villalobos} \& {Helmi}}{2008}]{Villalobos2008}
{Villalobos} {\'A}.,  {Helmi} A.,  2008, \mn@doi [\mnras] {10.1111/j.1365-2966.2008.13979.x}, \href {https://ui.adsabs.harvard.edu/abs/2008MNRAS.391.1806V} {391, 1806}

\bibitem[\protect\citeauthoryear{{Weilbacher}, {Streicher}  \& {Palsa}}{{Weilbacher} et~al.}{2016}]{Weilbacher2016}
{Weilbacher} P.~M.,  {Streicher} O.,   {Palsa} R.,  2016, {MUSE-DRP: MUSE Data Reduction Pipeline}, Astrophysics Source Code Library, record ascl:1610.004

\bibitem[\protect\citeauthoryear{{Weilbacher} et~al.,}{{Weilbacher} et~al.}{2020}]{Weilbacher2020}
{Weilbacher} P.~M.,  et~al., 2020, \mn@doi [\aap] {10.1051/0004-6361/202037855}, \href {https://ui.adsabs.harvard.edu/abs/2020A&A...641A..28W} {641, A28}

\bibitem[\protect\citeauthoryear{{Wilkinson} et~al.,}{{Wilkinson} et~al.}{2015}]{Wilkinson2015}
{Wilkinson} D.~M.,  et~al., 2015, \mn@doi [\mnras] {10.1093/mnras/stv301}, \href {https://ui.adsabs.harvard.edu/abs/2015MNRAS.449..328W} {449, 328}

\bibitem[\protect\citeauthoryear{{Williams}, {Dalcanton}, {Dolphin}, {Holtzman}  \& {Sarajedini}}{{Williams} et~al.}{2009}]{Williams2009}
{Williams} B.~F.,  {Dalcanton} J.~J.,  {Dolphin} A.~E.,  {Holtzman} J.,   {Sarajedini} A.,  2009, \mn@doi [\apjl] {10.1088/0004-637X/695/1/L15}, \href {https://ui.adsabs.harvard.edu/abs/2009ApJ...695L..15W} {695, L15}

\bibitem[\protect\citeauthoryear{{Williams}, {Zamojski}, {Bureau}, {Kuntschner}, {Merrifield}, {de Zeeuw}  \& {Kuijken}}{{Williams} et~al.}{2011}]{Williams2011}
{Williams} M.~J.,  {Zamojski} M.~A.,  {Bureau} M.,  {Kuntschner} H.,  {Merrifield} M.~R.,  {de Zeeuw} P.~T.,   {Kuijken} K.,  2011, \mn@doi [\mnras] {10.1111/j.1365-2966.2011.18535.x}, \href {https://ui.adsabs.harvard.edu/abs/2011MNRAS.414.2163W} {414, 2163}

\bibitem[\protect\citeauthoryear{{Xiang} et~al.,}{{Xiang} et~al.}{2018}]{Xiang2018}
{Xiang} M.,  et~al., 2018, \mn@doi [\apjs] {10.3847/1538-4365/aad237}, \href {https://ui.adsabs.harvard.edu/abs/2018ApJS..237...33X} {237, 33}

\bibitem[\protect\citeauthoryear{{Yoachim} \& {Dalcanton}}{{Yoachim} \& {Dalcanton}}{2006}]{Yoachim2006}
{Yoachim} P.,  {Dalcanton} J.~J.,  2006, \mn@doi [\aj] {10.1086/497970}, \href {https://ui.adsabs.harvard.edu/abs/2006AJ....131..226Y} {131, 226}

\bibitem[\protect\citeauthoryear{{Yoachim} \& {Dalcanton}}{{Yoachim} \& {Dalcanton}}{2008}]{Yoachim2008b}
{Yoachim} P.,  {Dalcanton} J.~J.,  2008, \mn@doi [\apj] {10.1086/590246}, \href {https://ui.adsabs.harvard.edu/abs/2008ApJ...683..707Y} {683, 707}

\bibitem[\protect\citeauthoryear{{Yu} et~al.,}{{Yu} et~al.}{2021}]{Yu2021}
{Yu} S.,  et~al., 2021, \mn@doi [\mnras] {10.1093/mnras/stab1339}, \href {https://ui.adsabs.harvard.edu/abs/2021MNRAS.505..889Y} {505, 889}

\bibitem[\protect\citeauthoryear{{Yu} et~al.,}{{Yu} et~al.}{2023}]{Yu2023}
{Yu} S.,  et~al., 2023, \mn@doi [\mnras] {10.1093/mnras/stad1806}, \href {https://ui.adsabs.harvard.edu/abs/2023MNRAS.523.6220Y} {523, 6220}

\bibitem[\protect\citeauthoryear{{Yurin} \& {Springel}}{{Yurin} \& {Springel}}{2015}]{Yurin2015}
{Yurin} D.,  {Springel} V.,  2015, \mn@doi [\mnras] {10.1093/mnras/stv1454}, \href {https://ui.adsabs.harvard.edu/abs/2015MNRAS.452.2367Y} {452, 2367}

\bibitem[\protect\citeauthoryear{{Zabel} et~al.,}{{Zabel} et~al.}{2019}]{Zabel2019}
{Zabel} N.,  et~al., 2019, \mn@doi [\mnras] {10.1093/mnras/sty3234}, \href {https://ui.adsabs.harvard.edu/abs/2019MNRAS.483.2251Z} {483, 2251}

\bibitem[\protect\citeauthoryear{{Zibetti}, {Rossi}  \& {Gallazzi}}{{Zibetti} et~al.}{2024}]{Zibetti2024}
{Zibetti} S.,  {Rossi} E.,   {Gallazzi} A.~R.,  2024, \mn@doi [\mnras] {10.1093/mnras/stae178}, \href {https://ui.adsabs.harvard.edu/abs/2024MNRAS.528.2790Z} {528, 2790}

\bibitem[\protect\citeauthoryear{{de Vaucouleurs}, {de Vaucouleurs}, {Corwin}, {Buta}, {Paturel}  \& {Fouque}}{{de Vaucouleurs} et~al.}{1991}]{deVaucouleurs1991}
{de Vaucouleurs} G.,  {de Vaucouleurs} A.,  {Corwin} Jr. H.~G.,  {Buta} R.~J.,  {Paturel} G.,   {Fouque} P.,  1991, {Third Reference Catalogue of Bright Galaxies}

\bibitem[\protect\citeauthoryear{{van Donkelaar}, {Agertz}  \& {Renaud}}{{van Donkelaar} et~al.}{2022}]{vanDonkelaar2022}
{van Donkelaar} F.,  {Agertz} O.,   {Renaud} F.,  2022, \mn@doi [\mnras] {10.1093/mnras/stac692}, \href {https://ui.adsabs.harvard.edu/abs/2022MNRAS.512.3806V} {512, 3806}

\bibitem[\protect\citeauthoryear{{van de Sande}, {Fraser-McKelvie}, {Fisher}, {Martig}, {Hayden}  \& {Geckos Survey Collaboration}}{{van de Sande} et~al.}{2024}]{vandeSande2024}
{van de Sande} J.,  {Fraser-McKelvie} A.,  {Fisher} D.~B.,  {Martig} M.,  {Hayden} M.~R.,   {Geckos Survey Collaboration} 2024, in {Tabatabaei} F.,  {Barbuy} B.,   {Ting} Y.-S.,  eds,  IAU Symposium Vol. 377, Early Disk-Galaxy Formation from JWST to the Milky Way. pp 27--33 (\mn@eprint {arXiv} {2306.00059}), \mn@doi{10.1017/S1743921323001138}

\bibitem[\protect\citeauthoryear{{van der Kruit}}{{van der Kruit}}{1988}]{vanderKruit1988}
{van der Kruit} P.~C.,  1988, \aap, \href {https://ui.adsabs.harvard.edu/abs/1988A&A...192..117V} {192, 117}

\bibitem[\protect\citeauthoryear{{van der Kruit} \& {Searle}}{{van der Kruit} \& {Searle}}{1981}]{vanderKruit1981}
{van der Kruit} P.~C.,  {Searle} L.,  1981, \aap, \href {https://ui.adsabs.harvard.edu/abs/1981A&A....95..105V} {95, 105}

\makeatother
\end{thebibliography}

\
\appendix
\section{Density maps of all mono-age populations}\label{sec:appendix_maps}

\begin{figure*}
\includegraphics[width=\textwidth]{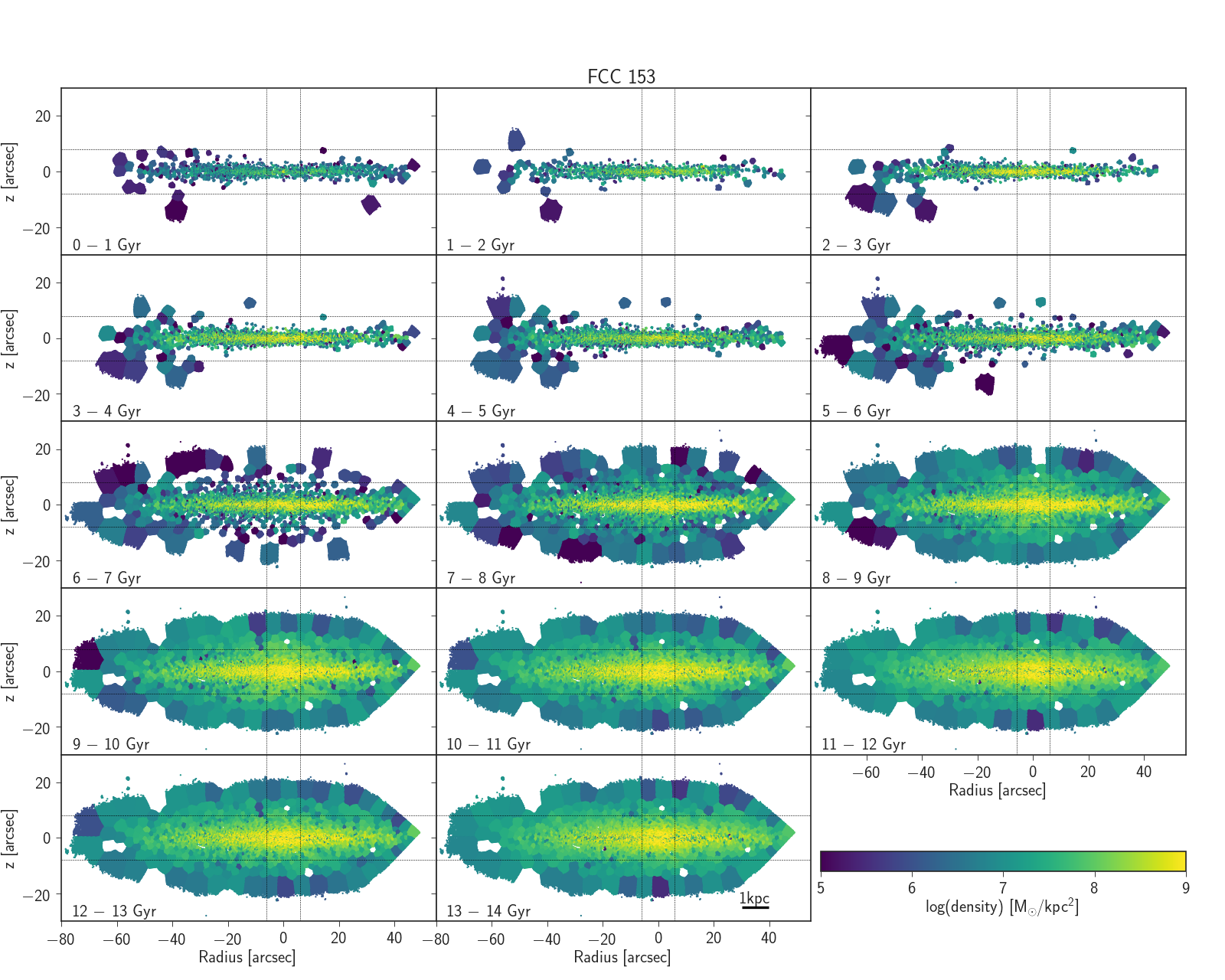}
\caption[]{This figure shows the surface density for all mono-age populations in FCC 153 (from 0--1 Gyr old in the top left panel, to 13--14 Gyr old in the bottom right panel). In all panels, the vertical lines enclose the central bulge or bar-dominated region, while the horizontal lines delimitate the thin- and thick disc-dominated regions \citep{Pinna2019a,Pinna2019b}.} 
\label{fig:maps_FCC153}
\end{figure*}

\begin{figure*}
\includegraphics[width=\textwidth]{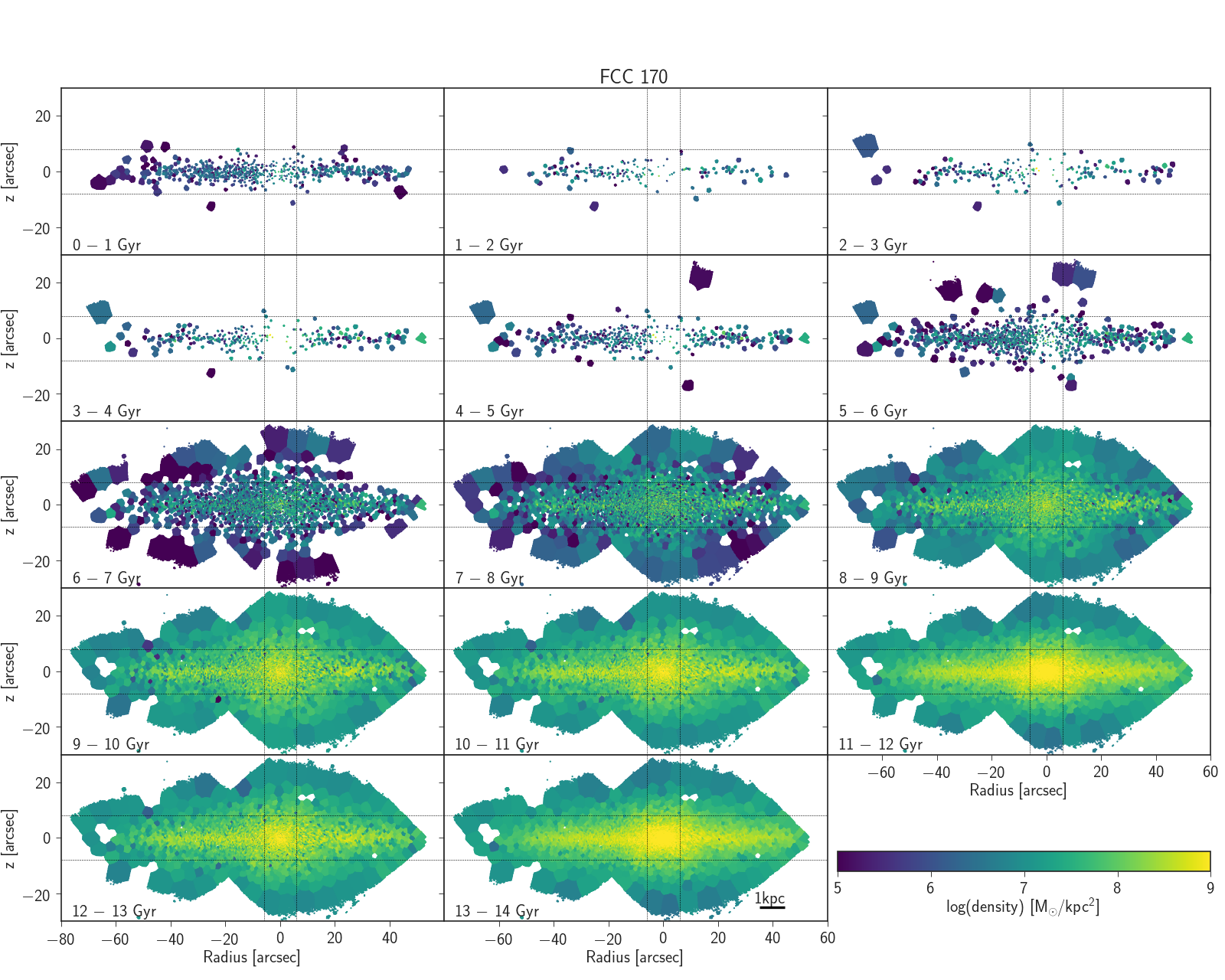}
\caption[]{This figure shows the surface density for all mono-age populations in FCC 170 (from 0--1 Gyr old in the top left panel, to 13--14 Gyr old in the bottom right panel). In all panels, the vertical lines enclose the central bulge or bar-dominated region, while the horizontal lines delimitate the thin- and thick disc-dominated regions \citep{Pinna2019a,Pinna2019b}.} 
\label{fig:maps_FCC170}
\end{figure*}

\begin{figure*}
\includegraphics[width=\textwidth]{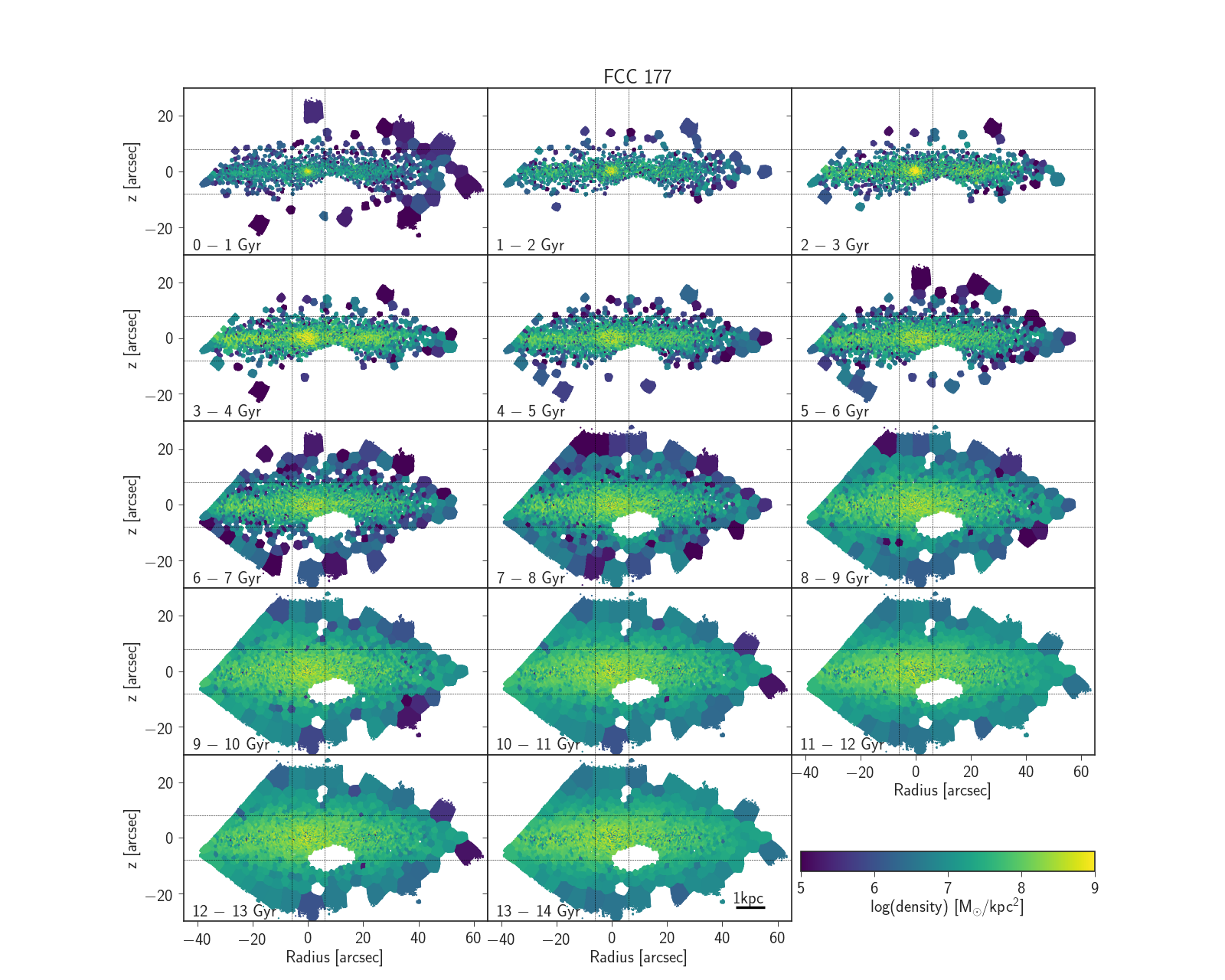}
\caption[]{This figure shows the surface density for all mono-age populations in FCC 177 (from 0--1 Gyr old in the top left panel, to 13--14 Gyr old in the bottom right panel). In all panels, the vertical lines enclose the central bulge or bar-dominated region, while the horizontal lines delimitate the thin- and thick disc-dominated regions \citep{Pinna2019a,Pinna2019b}.} 
\label{fig:maps_FCC177}
\end{figure*}

In this appendix, we show the surface density maps for all mono-age populations in FCC 153 (Fig. \ref{fig:maps_FCC153}), in FCC 170 (Fig. \ref{fig:maps_FCC170}) and in FCC 177 (Fig. \ref{fig:maps_FCC177}).

\bsp	
\label{lastpage}
\end{document}